\documentclass{article} 

\usepackage[figuresright]{rotating}
\usepackage[utf8]{inputenc}
\usepackage{amsmath}
\usepackage{tablefootnote}
\usepackage{booktabs}
\usepackage{subfigure}
\usepackage{dblfloatfix}
\usepackage{float}
\usepackage{graphicx}
\usepackage{authblk}

\usepackage{multirow}
\usepackage{array} 
\usepackage[hyphens]{url}
\usepackage{hyperref}
\usepackage[margin=1in]{geometry}
\usepackage{enumerate}

\DeclareMathOperator*{\argmin}{argmin}

\title{Personalized Biopsy Schedules Using an Interval-censored Cause-specific Joint Model}

\author[1,2]{Zhenwei Yang}
\author[1,2]{Dimitris Rizopoulos}
\author[3]{Eveline A.M. Heijnsdijk}
\author[4]{Lisa F. Newcomb}
\author[1,2]{Nicole S. Erler}
\affil[1]{Department of Biostatistics, Erasmus Medical Center Rotterdam}
\affil[2]{Department of Epidemiology, Erasmus Medical Center Rotterdam}
\affil[3]{Department of Public Health, Erasmus Medical Center Rotterdam}
\affil[4]{Fred Hutchinson Cancer Center, Cancer Prevention Program, Public Health Sciences, Seattle, Washington}

\begin{document}


\maketitle

\begin{abstract}
Active surveillance (AS), where biopsies are conducted to detect cancer progression, has been acknowledged as an efficient way to reduce the overtreatment of prostate cancer. Most AS cohorts use fixed biopsy schedules for all patients. However, the ideal test frequency remains unknown, and the routine use of such invasive tests burdens the patients. An emerging idea is to generate personalized biopsy schedules based on each patient's progression-specific risk. To achieve that, we propose the interval-censored cause-specific joint model (ICJM), which models the impact of longitudinal biomarkers on cancer progression while considering the competing event of early treatment initiation. The underlying likelihood function incorporates the interval-censoring of cancer progression, the competing risk of treatment, and the uncertainty about whether cancer progression occurred since the last biopsy in patients that are right-censored or experience the competing event. The model can produce patient-specific risk profiles until a horizon time. If the risk exceeds a certain threshold, a biopsy is conducted. The optimal threshold can be chosen by balancing two indicators of the biopsy schedules: the expected number of biopsies and expected delay in detection of cancer progression. A simulation study showed that our personalized schedules could considerably reduce the number of biopsies per patient by 41\%-52\% compared to the fixed schedules, though at the cost of a slightly longer detection delay.
\end{abstract}


\section{Introduction}
Prostate cancer is the most frequently diagnosed and the second most commonly occurring cancer in men worldwide.\cite{GANDAGLIA2021877} However, most cases are indolent and immediate treatment is usually unnecessary.\cite{Pernar2018, Albertsen2005} An analysis of the Medicare and Surveillance, Epidemiology and End Results (SEER) dataset estimated an overtreatment rate of 67\% in low-risk patients with prostate cancer.\cite{aizer2015} Since active surveillance (AS) has been widely acknowledged as an efficient way to reduce overtreatment,\cite{Chen2016} a consensus has been reached that low-risk patients should enter AS and defer treatment until the confirmation of cancer progression. Information on the tumor status (and, hence, whether the cancer has progressed) can be inferred from biomarkers such as the prostate specific antigen (PSA) as well as measurements obtained during biopsies.

There exist several AS protocols implemented in different countries. The majority of the existing AS cohorts used fixed biopsy schedules for all patients, with frequencies ranging from every one to four years.\cite{Bul2013, Tosoian2011, Klotz2015, Welty2015, Adamy2011, Davis2016, Newcomb2016, Godtman2016, Selvadurai2013, Lowenstein2019} However, due to the lack of evidence for the ideal test interval, the discussion about biopsy schedules is still ongoing.\cite{Bruinsma2016} Dall'Era et al.\cite{Marc2008} reported a cancer progression rate of 31\% in AS, indicating that most patients regularly undergo unnecessary examinations. A particular difficulty in determining an appropriate frequency for regular biopsies is that frequent biopsies are associated with an increased patient burden and risk for complications. In contrast, infrequent biopsies increase the chance of cancer progression being missed, resulting in more adverse clinical outcomes. Biopsy schedules that safely relieve the burden of frequent biopsies and balance the harms and benefits are needed. To that effect, we argue that biopsy schedules should be customized based on the individuals' risk. 

Although such personalized AS schedules are not yet implemented in prostate cancer, different studies have proposed methodologies for individualized cancer surveillance. Zhou et al.\cite{Zhou2020} elaborated on selecting the nasopharyngeal carcinoma posttreatment surveillance strategies from a pool of schedule candidates constructed according to the time-specific occurrence probabilities for the time to disease recurrence derived from a random survival forest. They selected the optimal personalized schedule as the most cost-effective one in the simulation from a Markov model. However, their model only distinguished between different risk groups and not individual patients. Moreover, the model described by Zhou et al.\cite{Zhou2020} can handle right censored survival data,\cite{Hemant2008} whereas an additional challenge in cancer AS is that, due to the periodical examinations, the time of progression is not observed exactly, resulting in interval censoring. Despite the limited literature discussing ideas of scheduling, recent studies focus on the risk models that can serve for the decision-making process.\cite{Mamawala2017, Madhur2022, Eminaga2022} One recent paper employed a deep learning technique, the Dynamic-DeepHit-Lite model, for determining the individual's risk of cancer progression.\cite{Lee2022} The model also included the longitudinal trajectory of the PSA, which is believed to be a sensitive biomarker for prostate cancer.\cite{Cary2013} Generally, in the observed data, repeated biomarker measurements require special caution since they are typically endogenous and likely affected by measurement error.


So far, only Tomer et al.\cite{Ani2019, Ani2022} proposed a methodology for prostate cancer AS that can handle the biomarkers' dynamic nature, endogeneity, and measurement error, as well as the interval-censoring of the event of interest. The authors used a joint model for longitudinal and time-to-event data, in which the risk of cancer progression was derived from each patient's trajectories of the biomarkers over time, and is thus patient-specific. Nonetheless, these authors did not consider the informative censoring due to an early initiation of treatment. Typically a patient's decision to leave AS to seek treatment is associated with his (perceived) risk of progression and initiating treatment changes this risk. In the Canary Prostate Active Surveillance Study (PASS) that motivates our work, around 10\% of the participants initiated treatment before cancer progression had been detected.\cite{Cooperberg2020} The failure to account for the competing risks can bias the estimated cumulative risk of progression and eventually mislead the decision-making for personalized schedules.\cite{Schuster2020} A popular alternative methodology to suggest treatment options based on the patient's situation at different stages is Dynamic Treatment Regimes (DTR).\cite{Chakraborty2014, Tsiatis2020} DTR work in discrete time \cite{Hua2021} whereas the joint model predicts the risk in continuous time which provides more flexibility for determining the patient-specific optimal time for a clinical intervention or diagnostic test.


Our current study extends the model and theoretical framework for personalized biopsy schedules proposed by Tomer et al.\cite{Ani2022} to a setting with both interval-censoring and competing risks. This scenario is common in many medical applications, for example, when the diagnosis of presence or progression of a disease requires a (periodic) test and death is often a competing event in most of the cases. Specifically in the Canary PASS data, different events may have different censoring patterns (i.e., the time of progression is interval censored; the time of treatment initiation is exactly observed). Consequently, the censoring times may differ per event type: while the time to treatment initiation is censored when a patient leaves AS for another reason, the time to cancer progression for the same patient is censored at the last available biopsy. To overcome this challenge while also considering the aforementioned dynamic nature of (endogenous) biomarkers that may be measured with error, we develop an interval-censored cause-specific joint model (ICJM) that combines a cause-specific proportional hazard model with a multivariate mixed model. Our model incorporates the ``core ratio" as an additional biomarker (besides PSA) to improve the precision of the risk of progression. The ``core ratio" denotes the proportion of samples taken during a biopsy that contain cancer cells. A particular feature that needs consideration is that since the core ratio is obtained from a biopsy, measurement times generally differ from the typically more frequent times at which PSA measurements are taken.

Taking advantage of the ICJM in the context of personalized scheduling, biopsies are scheduled when the estimated individual progression-specific risk, derived from the ICJM, exceeds a certain threshold. Under the assumption that the scheduled biopsy will indicate absence of progression, the risk is reset to zero at that time, and the subsequent biopsy can be planned in the same manner. Different choices of thresholds will result in different numbers of planned biopsies (fewer seen as benefits) and different detection delays (longer seen as harm). To further tailor the schedules to each patient's risk profile, we determine these time-dependent and patient-specific thresholds by balancing the two indicators for each patient. Schedules can be dynamically updated whenever additional information on the biomarkers becomes available. The source codes of the ICJM and personalized biopsy schedules can be found on Github (\url{https://github.com/ZhenweiYang96/ICJM_Precision_Medicine}).

The remainder of this paper is organized as follows. In Section~\ref{sec:data}, we describe the motivating Canary PASS data. Section~\ref{sec:jm} defines the proposed interval-censored cause-specific joint model, and the personalized scheduling strategy is elaborated in Section~\ref{sec:biopsymeth}. The results from two ICJMs fitted on Canary PASS data are presented in Section~\ref{sec:analysis}. Section~\ref{sec:simulation} presents a simulation study to compare our personalized schedules with the existing fixed schedules for biopsies in AS. Section~\ref{sec:discussion} concludes this paper with a discussion.

\section{Canary PASS Data} \label{sec:data}

The Canary PASS trial is an ongoing multicenter active surveillance study in the United States.\cite{Newcomb2016} Patients are monitored with PSA tests every three months, clinical visits every six months, and biopsies at 12, 24 months, and biennially afterwards. The biopsies determine the tumor status, which is reflected in the Gleason score. In 2017, the trial included 850 patients diagnosed since 2003 with a Gleason score of six, who were found with a Gleason score of six or no cancer on confirmatory biopsy.\cite{Cooperberg2020} Our current study includes 833 patients with at least one PSA value recorded while on AS. The primary outcome, cancer progression, is defined as a Gleason score of seven or higher. Patients who reach this endpoint leave AS to receive treatment. Notably, 87 patients left AS for treatment before cancer progression was detected, constituting the competing event. A summary of the data is presented in Web Appendix 1.1. The Aalen-Johansen estimator for the risks of the two events, progression and early treatment, is visualized in Web Appendix 1.2.

Previous research has shown that the level of PSA, as well as the velocity of PSA level change, are predictive of cancer progression.\cite{Nelson2021, Cary2013} Knowledge about this association may 
induce patients to start early treatment when they see increasing PSA values. Hence, our aim here is to model the relation between PSA and cancer progression while considering the competing risk and to obtain risk estimates from this model that allow us to generate risk-based biopsy schedules. Furthermore, we also explore another clinically relevant biomarker, the core ratio, as an additional time-varying predictor variable in the risk model. The two longitudinal outcomes are visualized for a random subset of patients in Web Appendix 1.2.

\section{Interval-censored Cause-specific Joint Models (ICJM) for Longitudinal and Time-to-event Data} \label{sec:jm}

To link the repeated measurements of the PSA and core ratio with the time to cancer progression (or early treatment initiation), the ICJM jointly models a longitudinal component, in which the underlying development of the biomarkers is captured, and a time-to-event component, in which the risks of the primary and competing events are modeled. Specifically for the Canary PASS data, let $T^*_i$ be the true time of the event for the $i$-th patient, either the true time of cancer progression, or the time of early treatment initiation, $T^*_i \in \{T_i^{\textsc{prg}*}, T_i^{\textsc{trt}*}\}$, which happens first. The censoring time is denoted by $T^{\textsc{cen}}_i$. We consider two types of censoring, interval and right censoring. With interval censoring involved, the event-free time in a one-event setting becomes more complex, since the progression-free time determined by the biopsies may generally differ from the treatment-free time. For patients recorded with the interval-censored event, cancer progression ($\delta_i=1$, i.e., $T_i^{\textsc{prg}*}<\min\left\{T^{\textsc{cen}}_i, T_i^{\textsc{trt}*}\right\}$), the true event time is located between the last progression-free biopsy (i.e., the observed progression-free time), $T^{\textsc{prg-}}_i$, and the biopsy time at which progression was detected, $T^{\textsc{prg+}}_i$. Since those patients did not initiate early treatment, their treatment-free time is the time at which progression was detected, i.e., $T^{\textsc{prg+}}_i$.  For the observed event, initiation of treatment ($\delta_i=2$, i.e., $T_i^{\textsc{trt}*}<\min\left\{T^{\textsc{cen}}_i ,T_i^{\textsc{prg}*}\right\}$), the true event time $T_i^{\textsc{trt}*}$ equals the observed treatment-free time as well as the observed early treatment time $T^{\textsc{trt}}_i$  since the event is exactly observed. Note that the progression-free time ($T^{\textsc{prg-}}_i$), in this case, is not the treatment-free time, but the last biopsy time. When a patient is right-censored ($\delta_i=0$, i.e., $T^{\textsc{cen}}_i<\min\left\{T_i^{\textsc{prg}*}, T_i^{\textsc{trt}*}\right\}$), e.g., at the time the data was extracted from the database or because the patient was lost-to-follow-up, the progression-free time ($T^{\textsc{prg-}}_i$), is the time of the last recorded biopsy, while the treatment-free time is the censoring time ($T^{\textsc{cen}}_i$). In summary, the vector of the observed times for patient $i$ is given by
\begin{align*}
\boldsymbol{T}_i = 
    \begin{cases}
    \left[T^{\textsc{prg-}}_i, T^{\textsc{cen}}_i\right], &\text{if }\delta_i=0, \\
    \left[T^{\textsc{prg-}}_i, T^{\textsc{prg+}}_i\right], &\text{if }\delta_i=1, \\
    \left[T^{\textsc{prg-}}_i, T^{\textsc{trt}}_i\right], &\text{if }\delta_i=2.
    \end{cases}
\end{align*}

We denote the repeated observations of the two longitudinal outcomes, PSA level and core ratio, with $\boldsymbol{y}_{\textsc{psa}, i}$ and $\boldsymbol{y}_{\textsc{cr},i}$ for the $i$-th patient, respectively. The observed data, denoted by $\boldsymbol{\mathcal{D}}_n = \{\boldsymbol{T}_i, \boldsymbol{y}_{\textsc{psa}, i}, \boldsymbol{y}_{\textsc{cr},i}, \boldsymbol{x}_i; i = 1, \dots, n\}$ includes information on $n$ subjects, where $\boldsymbol{x}_i$ stands for the covariate information, such as age at start of AS, PSA density, biomarker measurement times (described in details in the following two sections).

\subsection{Longitudinal Component}

To capture the development of the longitudinal outcomes over time, generalized linear mixed models are adopted. Mixed models have the advantage that they allow us to consider multiple longitudinal outcomes of different types which may be measured at different time points as is the case for the PSA and core ratio.

PSA values are typically right-skewed and previous work has suggested to model the logarithm of $\text{PSA} + 1$.\cite{Pearson1994, Whittemore1995} Moreover, preliminary investigations in our own data suggested the assumption of a Student's $t$-distribution (with three degrees of freedom) for the error terms, $\epsilon_i(t)$. To capture the non-linear evolution of the PSA trajectories, we used a natural cubic spline (with three degrees of freedom) for the effect of time in the fixed and random effects ($\boldsymbol{u}$). The corresponding knots were placed according to the quantiles of the longitudinal measurement times and event times. The model for PSA, hence, had the following structure:
\begin{equation}
\begin{aligned}[b]
    \log_2\{\text{PSA}_i(t) + 1\} &= m_{\textsc{psa},i}(t) + \epsilon_i(t),  \\
   m_{\textsc{psa},i}(t) &= \beta_0 + u_{0i} + \sum^3_{p=1}(\beta_p + u_{pi})\mathcal{C}^{(p)}_i(t) + \beta_4(\text{Age}_i - 62),
\end{aligned}
\label{eq:psa}
\end{equation}
where $\mathcal{C}(t)$ is the design matrix for the natural cubic splines, and $\text{Age}_i$ refers to the patient's age at the start of active surveillance centered by subtracting the median age (62 years) for computational reasons.

The other repeatedly measured outcome, the core ratio, is the proportion of cores containing cancerous cells. Due to different biopsies, patients had different numbers of cores sampled (assumed either 12 or 100) at different follow-up times. An exploratory analysis showed a nonlinear trend in the core ratio over time. Compared to the PSA values, however, the core ratio has fewer measurements. We, thus, specified a binomial mixed model with a quadratic effect over time
\begin{equation}
\begin{aligned}[b]
    \text{logit}\left[E\left\{\text{core ratio}_i(t)\right\}\right] &= m_{\textsc{cr},i}(t), \\
    m_{\textsc{cr},i}(t) &= \beta_5 + u_{4i} + (\beta_6 + u_{5i})t + (\beta_7 + u_{6i})t^2,
\end{aligned}
\label{eq:cr}
\end{equation}
where $\text{logit}\left[E\left\{\text{core ratio}_i(t)\right\}\right] = \log \frac{\Pr\{\text{core ratio}_i(t) = 1\}}{1 - \Pr\{\text{core ratio}_i(t)=1\}}$. The random effects from the two longitudinal outcomes, $\boldsymbol{u}_i = (u_{0i}, \dots, u_{6i})^\top$, are modeled jointly using a multivariate normal distribution with mean zero and variance-covariance matrix $\boldsymbol{\Omega}$. Connecting the two longitudinal outcomes via their random effects allows us to jointly model them despite their different measurement times.

\subsection{Time-to-Event Component}

The longitudinal component is incorporated in a cause-specific proportional hazard model, which may include additional covariates. The model estimates the hazards of the competing events with separate parameters and baseline hazard functions. The hazard of patient $i$ to experience event $k \in \boldsymbol{\mathcal{K}}= \left\{\textsc{prg},\textsc{trt} \right\}$ at time $t$ is defined as:

\begin{align*}
    h_i^{(k)}\left\{t \mid \boldsymbol{\mathcal{M}}_{i}(t), \boldsymbol{w}_i(t)\right\} &= \lim_{\Delta t \rightarrow 0} \frac{\Pr\left\{t \leq T_i^{k*} < t + \Delta t \mid T^*_i \geq t, \boldsymbol{\mathcal{M}}_{i}(t), \boldsymbol{w}_i(t)\right\}}{\Delta t} \\
    &= h_0^{(k)}(t)\exp\left[\boldsymbol{\gamma}_k^\top\boldsymbol{w}_i(t) + \sum_{k\in \mathcal{K}}\sum_{q\in\mathcal{Q}}f_{kq}\{\boldsymbol{m}_{qi}(t),  \boldsymbol{\alpha}_{kq}\}\right],
\end{align*}
where $\boldsymbol{\mathcal{M}}_{i}(t) = \{\boldsymbol{m}_{\textsc{psa},i}(z), \boldsymbol{m}_{\textsc{cr},i}(z); 0 \leq z < t\}$ are the estimated trajectories of the longitudinal outcomes until time $t$, and $\boldsymbol{w}_i(t)$ is a vector of potentially time-varying exogenous covariates (in our case, the baseline PSA density), with corresponding regression coefficients $\boldsymbol{\gamma}_k$, $f_{kq}[\cdot]$ denotes the functional form for the effect of the longitudinal outcome $q \in \boldsymbol{\mathcal{Q}} = \left\{\textsc{psa},\textsc{cr} \right\}$, $\boldsymbol{y}_q$, on the hazard of event $k$, and $\boldsymbol{\alpha}_{kq}$ is the corresponding parameter vector. 

For the Canary PASS data, we explored two association structures for the PSA level, the effect of the expected value at time $t$ ($\log_2(\text{PSA} + 1)$ value) and the expected average change between times $t-1$ and $t$ ($\log_2(\text{PSA} + 1)$ yearly change), i.e.,
\begin{align*}
    f_{k,\textsc{psa}}\left\{\boldsymbol{\mathcal{M}}_{\textsc{psa},i}(t), \boldsymbol{\alpha}_{k,\textsc{psa}}\right\} &= \alpha_{1k,\textsc{psa}} m_{\textsc{psa},i}(t) + \alpha_{2k,\textsc{psa}}\left\{m_{\textsc{psa},i}(t) - m_{\textsc{psa},i}(t - 1)\right\},
\end{align*}
where with regard to the time points earlier than year one (i.e., $t<1$), extrapolation was employed based on the longitudinal part of the ICJM. For the core ratio, the expected value was considered
\begin{align*}
    f_{k,\textsc{cr}}\left\{\boldsymbol{\mathcal{M}}_{\textsc{cr},i}(t), \boldsymbol{\alpha}_{k,\textsc{cr}}\right\} = \alpha_{1k,\textsc{cr}} m_{\textsc{cr},i}(t).
\end{align*}

Contrary to previous studies where the velocity of a biomarker was represented as the ``current slope", mathematically denoted as $m'_{\textsc{psa},i}(t) = \frac{dm_{\textsc{psa},i}(t)}{dt}$, we opted for the yearly change because it is more straightforward to interpret from a clinical point of view.

\subsection{Estimation}

The ICJM can be estimated under the Bayesian framework using Markov chain Monte Carlo (MCMC) methods. The likelihood of the model is derived under the assumption that, conditional on the random effects $\boldsymbol{u}_{i}$, the survival and longitudinal part are independent, and the repeated measurements of the longitudinal outcomes for the same patient are independent. The likelihood for patient $i$ can be written as:

\begin{align*}
    p\left\{\boldsymbol{y}_{\textsc{psa},i}, \boldsymbol{y}_{\textsc{cr},i}, \boldsymbol{T}_i, \delta_i \mid \boldsymbol{u}_i, \boldsymbol{\theta}\right\} = \prod_{q\in\boldsymbol{\mathcal{Q}}} \prod^{n_{qi}}_{j=1}p(y_{qij} \mid \boldsymbol{u}_{qi}, \boldsymbol{\theta}) \times p\left\{\boldsymbol{T}_i, \delta_i \mid \boldsymbol{u}_i, \boldsymbol{\theta}\right\},
\end{align*}
where $\delta_i$ is the indicator of the censoring type with $\delta_i = 0$ for right-censored observations, $\delta_i = 1$ for cancer progression and $\delta_i = 2$ for early treatment initiation. $\boldsymbol{\theta}$ denotes the vector of all model parameters, i.e., $\boldsymbol{\beta}_q$, $\boldsymbol{\gamma}_k$ and $\boldsymbol{\alpha}_{kq}$, for $k \in \{\textsc{prg}, \textsc{trt}\}$ and $q \in \{\textsc{psa}, \textsc{cr}\}$. $n_{qi}$ is the number of repeat measurements of the longitudinal outcome $q$ for patient $i$. The first term of the likelihood formula stands for the likelihood of the longitudinal outcomes conditional on the random effects and parameters. Taking into account the combination of multiple competing events that may be right- or interval-censored, the likelihood of the time-to-event component for patient $i$ can be written as:

\begin{align*}
    p(\boldsymbol{T}_i, \delta_i \mid \boldsymbol{u}_i, \boldsymbol{\theta}) &= \left[\exp\left\{-
    \int^{T_i^{\textsc{prg-}}}_0 h^{(\textsc{prg})}_i(\nu)d\nu - \int^{T_i^{\textsc{cen}}}_0 h^{(\textsc{trt})}_i(\nu)d\nu\right\}\right]^{I(\delta_i = 0)} \\  
    &\ \  \times \Bigg[\int^{T^{\textsc{prg+}}_i}_{T^{\textsc{prg-}}_i}h^{(\textsc{prg})}_i(s) \exp\Bigg\{- \int^s_0h^{(\textsc{prg})}_i(\nu)d\nu  - \int^{T^{\textsc{prg+}}_i}_0 h^{(\textsc{trt})}_i(\nu)d\nu\Bigg\}ds\Bigg]^{I(\delta_i = 1)} \\
    &\ \  \times \Bigg[h^{(\textsc{trt})}_i(T^{\textsc{trt}}_i) \exp\Bigg\{-\int^{T_i^{\textsc{prg-}}}_0h^{(\textsc{prg})}_i(\nu)d\nu -\int^{T^{\textsc{trt}}_i}_0 h^{(\textsc{trt})}_i(\nu)d\nu\Bigg\}\Bigg]^{I(\delta_i = 2)}.
\end{align*}
The first factor (for $\delta_i = 0$), is the probability of not having experienced any event up until the respective event-free times. Since it is only known that event $k$ did not happen until the last available observation of the event status, 
patients contribute to the ``overall survival" part of the likelihood only until their event-specific event-free times, the last biopsy time $T^{\textsc{prg-}}_i$ and censoring time ($T^{\textsc{cen}}_i$). The second factor (for $\delta_i = 1$) models the probability of patients having progression between the interval between the last progression-free biopsy, $T^{\textsc{prg-}}_i$, and the biopsy at which progression was detected, $T^{\textsc{prg+}}_i$. For those patients the ``overall survival" part includes the probability that patients are progression-free until time $s$, where $s$ ranges over the interval $(T^{\textsc{prg-}}_i, T^{\textsc{prg+}}]$, as well as the probability that the patient did not initiate early treatment before the detection of progression, i.e., until $T^{\textsc{prg+}}_i$. The third factor (for $\delta_i = 2$) uses the standard cause-specific cumulative incidence function in the ``overall survival" part to formulate the probability of patients starting early treatment at $T^{\textsc{trt+}}_i$, conditional on patients not having progressed until the last biopsy (i.e., progression-free time) $T^{\textsc{prg-}}_i$. The integrals in the above equation do not have a closed-form solution, and can be numerically approximated using the 15-point Gauss-Kronrod rule. Model estimation using Gibbs sampling combined with the Metropolis–Hastings algorithm can be easily implemented in available software, such as JAGS.\cite{JAGS}

\section{Personalized Biopsy Schedules} \label{sec:biopsymeth}


\subsection{Personalized Risk of Progression} \label{prisk}

The ICJM utilizes the full history of relevant biomarker measurements for any patient and models the patient-specific trajectories of these markers over time. This enables us to obtain patient-specific and dynamic risk estimates for the main event of interest, cancer progression. Key to this is that the patient-specific information from the longitudinal outcomes is captured in the random effects, which link the longitudinal and survival part of the model. The patient-specific risk estimates for a new patient $i'$ can, thus, be obtained via the random effects derived from the history of the biomarker values for that patient.

Let $t^{(p)}$ be the time in the future at which the risk of patient $i'$ should be predicted, $t^{(b)}$ the time at which the most recent biopsy was taken (i.e., until which the cancer progression is assured not to have happened), $t_v$ is the current $v$-th clinical visit among $V$ planned visits after the start of AS (i.e., until when the early treatment is assured not to have happened), $t^{(y)}$ the time of the latest measurement of the longitudinal outcomes, either PSA or the core ratio. The current visit time, $t_v$, is typically between $t^{(b)}$ and $t^{(p)}$, $t^{(b)} \leq t_v \leq t^{(p)}$. Because monitoring patients on AS and scheduling biopsies are dynamic processes, and thus those four time points typically change over time. 

For brevity, the baseline covariates for patient $i'$, $\text{Age}_{i'}$ and $\boldsymbol{w}_{i'}(t)$, are omitted in the following formulas. Using the history of the longitudinal outcomes until $t^{(y)}$, we can formulate the progression-specific cumulative risk at time $t^{(p)}$

\begin{equation}
\begin{aligned}[b]
    \Pi^{(\textsc{prg})}_{i'}\left\{t^{(p)} \mid t^{(b)}, t_v, t^{(y)}\right\} &= \Pr\Bigg[T^{\textsc{prg}*}_{i'} \leq t^{(p)}, T^{\textsc{prg}*}_{i'} < T^{\textsc{trt}*}_{i'} \mid T^{\textsc{prg}*}_{i'} > t^{(b)}, T^{\textsc{trt}*}_{i'} > t_v, \boldsymbol{\mathcal{Y}}_{i'}\left\{t^{(y)}\right\}, \boldsymbol{\mathcal{D}}_n\Bigg] \\
    &= \int \int \Pr\Big\{T^{\textsc{prg}*}_{i'} \leq t^{(p)}, T^{\textsc{prg}*}_{i'} < T^{\textsc{trt}*}_{i'} \mid T^{\textsc{prg}*}_{i'} > t^{(b)}, T^{\textsc{trt}*}_{i'} > t_v, \boldsymbol{u}_{i'}, \boldsymbol{\theta}\Big\} \\
    &\qquad\qquad \Pr\Big[\boldsymbol{u}_{i'} \mid T^{\textsc{prg}*}_{i'} > t^{(b)}, T^{\textsc{trt}*}_{i'} > t_v, \boldsymbol{\mathcal{Y}}_{i'}\left\{t^{(y)}\right\}, \boldsymbol{\theta}\Big] \\
    &\qquad\qquad \Pr(\boldsymbol{\theta} \mid \boldsymbol{\mathcal{D}}_n) d\boldsymbol{u}_{i'} d\boldsymbol{\theta},\ t^{(b)} \leq t_v \leq t^{(p)},
\end{aligned}
\label{eq:riskpred}
\end{equation}
where the first term inside the integral can be rewritten based on the Bayes rule as
\begin{equation}
\begin{aligned}[b]
    &\Pr\Big\{T^{\textsc{prg}*}_{i'} \leq t^{(p)}, T^{\textsc{prg}*}_{i'} < T^{\textsc{trt}*}_{i'} \mid T^{\textsc{prg}*}_{i'} > t^{(b)}, T^{\textsc{trt}*}_{i'} > t_v, \boldsymbol{u}_{i'}, \boldsymbol{\theta}\Big\} \\
    &=\frac{\Pr\left\{t^{(b)}<T^{\textsc{prg}*}_{i'} \leq t^{(p)}, \max(T^{\textsc{prg}*}_{i'},t_v) < T^{\textsc{trt}*}_{i'} \mid \boldsymbol{u}_{i'}, \boldsymbol{\theta}\right\}}{\Pr\left\{T^{\textsc{prg}*}_{i'} > t^{(b)}, T^{\textsc{trt}*}_{i'} > t_v\mid \boldsymbol{u}_{i'}, \boldsymbol{\theta}\right\}} \\
    &=\frac{\displaystyle{\int}^{t^{(p)}}_{t^{(b)}} h^{(\textsc{prg})}_{i'}(\nu)\exp\left[-H^{(\textsc{prg})}_{i'}(\nu)-H^{(\textsc{trt})}_{i'}\{\max(\nu,t_v)\}d\nu\right]}{\exp\left[-H^{(\textsc{prg})}_{i'}\{t^{(b)}\}-H^{(\textsc{trt})}_{i'}(t_v)\right]},
\end{aligned}
\label{eq:riskformula}
\end{equation}
where $H^{\textsc(prg)}_{i'}(\cdot)$ and $H^{\textsc(trt)}_{i'}(\cdot)$ denote the cumulative hazard of cancer progression and early treatment, respectively, until a certain time point. Note that due to the competing risk setting, cause-specific prediction has to be specified in terms of the cumulative incidence function. In the formulation of the progression-specific cumulative risk, the competing risk of early treatment initiation is taken into account by using the joint probability of progression happening before $t^{(p)}$ and treatment not being initiated before $t_v$ or $T^{\textsc{prg}*}_{i'}$, whichever (is expected to) happen(s) first.

The second term inside the integral of (\ref{eq:riskpred}) is the posterior distribution of the random effects $\boldsymbol{u}_{i'}$ conditional on the longitudinal outcomes until time $t^{(y)}$, the patient being progression-free until $t^{(b)}$ and treatment-free until $t_v$, and the third term is the posterior distribution of the model parameters $\boldsymbol{\theta}$ conditional on the observed data $\boldsymbol{\mathcal{D}}_n$. 

Inference can be performed using the following Monto Carlo sampling scheme with $\mathcal{L}$ iterations\cite{Rizopoulos2011}: 
\begin{enumerate}[I.]
    \setlength{\itemsep}{0pt}
    \item sample $\boldsymbol{\theta}^{(\ell)}$ from the posterior distribution $ p(\boldsymbol{\theta} \mid \boldsymbol{\mathcal{D}}_n)$, 
    \item  sample random effects, $\boldsymbol{u}_{i'}^{(\ell)}$ from their posterior distribution $\Pr\Big[\boldsymbol{u}_{i'} \mid T^{\textsc{prg}*}_{i'} > t^{(b)}, T^{\textsc{trt}*}_{i'} > t_v, \boldsymbol{\mathcal{Y}}_{i'}\left\{t^{(y)}\right\}, \allowbreak\boldsymbol{\theta}^{(\ell)}\Big]$, 
    \item calculate the progression-specific risk, $ \Pr\Big\{T^{\textsc{prg}*}_{i'} \leq t^{(p)}, T^{\textsc{prg}*}_{i'} < T^{\textsc{trt}*}_{i'} \mid T^{\textsc{prg}*}_{i'} > t^{(b)}, T^{\textsc{trt}*}_{i'} > t_v, \boldsymbol{u}_{i'}^{(\ell)},\allowbreak \boldsymbol{\theta}^{(\ell)}\Big\}$, where $\ell$ is the iteration index. The posterior distribution of the random effects in step (2) is given by
\end{enumerate}
\begin{equation}
\begin{aligned}[b] 
    &\Pr\Big[\boldsymbol{u}_{i'} \mid T^{\textsc{prg}*}_{i'} > t^{(b)}, T^{\textsc{trt}*}_{i'} > t_v, \boldsymbol{\mathcal{Y}}_{i'}\left\{t^{(y)}\right\}, \boldsymbol{\theta}\Big] \propto \\
    &\quad \Pr\left\{T^{\textsc{prg}*}_{i'} > t^{(b)}, T^{\textsc{trt}*}_{i'} > t_v \mid \boldsymbol{u}_{i'}, \boldsymbol{\theta}\right\} \Pr\left[\boldsymbol{\mathcal{Y}}_{i'}\left\{t^{(y)}\right\} \mid \boldsymbol{u}_{i'}, \boldsymbol{\theta}\right] \Pr(\boldsymbol{u}_{i'} \mid \boldsymbol{\theta}),
\end{aligned}
\label{eq:ranef}
\end{equation}
which does not have a closed form. The Metropolis–Hastings algorithm can be used for sampling the random effects for subject $i'$. Specifically, we apply the adaptive automatic scaling of the Metropolis–Hastings algorithm using the Robbins-Monro process, as proposed by Garthwaite et al.\cite{Garthwaite2016}.
 
\subsection{Generation of Personalized Schedules} \label{ps}

To facilitate clinical decision-making, we translate the patient-specific risk estimates into monitoring decisions, i.e., whether a biopsy should be performed at the current clinical visit, and generate a tentative schedule for future biopsies. A concrete example of the personalized schedules can be found in Web Appendix 5.

This personalized biopsy schedule aims to provide patients with the most suitable biopsy frequencies according to their individual risk of cancer progression. Since AS is in place to avoid over-treatment, starting treatment early is an undesirable behavior in this context. Therefore, in our application, schedules are created for the desired scenario in which the patient does not start treatment early. This means biopsies are scheduled based on the predicted progression-specific cumulative risk and the expected number of biopsies and detection delay (used for determining patient-specific risk thresholds, see below) are calculated under the assumption that the patient will not drop out, i.e., will not, be censored or initiate early treatment before a specified horizon time.

Assume $\{t_1, \dots, t_V\}$ are the $V$ planned and typically regular clinical visits after the start of AS at which biopsies could be performed, up until a horizon time $t_V$, when one biopsy is always conducted. At the current clinical visit $t_v$ ($v \in \{1, \dots, V\}$), the risk estimate of interest is always the cumulative progression-specific risk (\ref{eq:riskpred}) at $t_v$ (i.e., $t^{(p)}=t_v$) conditional on the time of the previous biopsy which was typically conducted before the current visit ($t^{(b)}$), the patient not yet initiating early treatment until the current visit ($t_v$) and the most recent longitudinal information ($t^{(y)}$). Within the context, (\ref{eq:riskformula}) can be simplified to
\begin{align} \label{eq:risk_notrt}
\begin{split}
    \frac{\displaystyle{\int}^{t^{(p)}}_{t^{(b)}} h^{(\textsc{prg})}_{i'}(\nu)\exp\left\{-H^{(\textsc{prg})}_{i'}(\nu)\right\}d\nu}{\exp\left[-H^{(\textsc{prg})}_{i'}\{t^{(b)}\}\right]}, \ \text{if } t^{(p)}=t_v,
\end{split}
\end{align}
which indicates that the hazard of early treatment does not play a role in the risk of cancer progression and thus is not necessarily considered anymore. The risk estimate in (\ref{eq:riskpred}) can be simpliy rewritten as $\Pi^{(\textsc{prg})}_{i'}\left\{t^{(p)} \mid t^{(b)}, t^{(y)}\right\}$. Correspondingly, the updating of random effects in (\ref{eq:ranef}) is now exempted from the current visit as well
\begin{equation}
\begin{aligned}[b]
    \Pr\Big[\boldsymbol{u}_{i'} \mid T^{\textsc{prg}*}_{i'} > t^{(b)}, \boldsymbol{\mathcal{Y}}_{i'}\left\{t^{(y)}\right\}, \boldsymbol{\theta}\Big] \propto \Pr\left\{T^{\textsc{prg}*}_{i'} > t^{(b)}, \mid \boldsymbol{u}_{i'}, \boldsymbol{\theta}\right\} \Pr\left[\boldsymbol{\mathcal{Y}}_{i'}\left\{t^{(y)}\right\} \mid \boldsymbol{u}_{i'}, \boldsymbol{\theta}\right] \Pr(\boldsymbol{u}_{i'} \mid \boldsymbol{\theta}).
\end{aligned}
\label{eq:ranef_notrt}
\end{equation}

If this risk estimate at a certain visit, $t_v$, exceeds a threshold $\phi$, i.e., $\Pi^{(\textsc{prg})}_{i'}\Big\{t_v \mid t^{(b)}, t^{(y)}\Big\} \geq \phi, \quad \phi \in [0,1]$, a biopsy is scheduled at time $t_v$. To plan subsequent biopsies, we assume that the biopsy scheduled at $t_v$ does not reveal cancer progression. The estimated risk profile then restarts from zero at $t_v$. For calculating the risk estimate for the next clinical visit $v+1$ is then recalculated conditional on progression happening after the previous (planned) biopsy, $t_v$ is now the time of the most recent biopsy (i.e., the risk of interest becomes $\Pi^{(\textsc{prg})}_{i'}\Big\{t_{v+1} \mid t_v, t^{(y)}\Big\}$). If the risk estimate at $t_v$ does not exceed $\phi$, no biopsy is scheduled at $t_v$ and the risk estimate at $t_{v+1}$ remains conditional on progression happening after $t^{(b)}$, where $t^{(b)}$ remains the same as during the calculation for the risk at $t_{v}$. The risk threshold $\phi$ can either be pre-specified or determined for each patient in a data-driven manner. 

During the planning phase from the current visit $t_v$ onwards, for any clinical visit $t_e$ ($e \in \{v+1, \dots, V\}$) in the future, the risk estimate of interest at the current visit (\ref{eq:risk_notrt}) is the cumulative risk of cancer progression only conditional on the patient being progression-free until the previous planned biopsy. This previous (tentative) biopsy time $\tilde{t}^{(b)}$ ($t^{(b)} \leq \tilde{t}^{(b)} < t_e$), however, dynamically changes each time a new biopsy is scheduled. Since the posterior distribution of the random effects (\ref{eq:ranef_notrt}) is now conditional on the time of the previous planned biopsy, $\tilde{t}^{(b)}$, this distribution changes each time a biopsy is scheduled. Re-sampling the random effects each time a biopsy is scheduled for this patient, leads to a computationally intense procedure. However, this complexity can be eased by reformulating the progression-specific cumulative risk so that it is conditional on the last indeed conducted biopsy at $t^{(b)}$ which remains fixed throughout the scheduling.

\begin{align*}
    \Pi^{(\textsc{prg})}_{i'}\left\{t_{e} \mid \tilde{t}^{(b)}, t^{(y)}\right\}&= \Pr\Bigg[T^{\textsc{prg}*}_{i'} \leq t_e \mid T^{\textsc{prg}*}_{i'} > \tilde{t}^{(b)}, \boldsymbol{\mathcal{Y}}_{i'}\Big\{t^{(y)}\Big\}, \boldsymbol{\mathcal{D}}_n\Bigg] \\
    &=\frac{\Pr\Bigg[\tilde{t}^{(b)}<T^{\textsc{prg}*}_{i'} \leq t_e \mid T^{\textsc{prg}*}_{i'} > t^{(b)}, \boldsymbol{\mathcal{Y}}_{i'}\Big\{t^{(y)}\Big\}, \boldsymbol{\mathcal{D}}_n\Bigg]}{\Pr\Bigg[T^{\textsc{prg}*}_{i'} > \tilde{t}^{(b)} \mid T^{\textsc{prg}*}_{i'} > t^{(b)}, \boldsymbol{\mathcal{Y}}_{i'}\Big\{t^{(y)}\Big\}, \boldsymbol{\mathcal{D}}_n\Bigg] }, 
\end{align*}
where the numerator can be expressed as the difference between the risk estimates of cancer progression at $\tilde{t}^{(b)}$ and $t_e$ derived from the risk curve calculated from the previous conducted biopsy, i.e., $\Pi^{(\textsc{prg})}_{i'}\left\{t_{e} \mid t^{(b)}, t^{(y)}\right\} - \Pi^{(\textsc{prg})}_{i'}\left\{\tilde{t}^{(b)} \mid t^{(b)}, t^{(y)}\right\}$. The dominator has the following formula
\begin{align*}
    \Pr\Bigg[T^{\textsc{prg}*}_{i'} > \tilde{t}^{(b)} \mid T^{\textsc{prg}*}_{i'} > t^{(b)}, \boldsymbol{\mathcal{Y}}_{i'}\Big\{t^{(y)}\Big\}, \boldsymbol{\mathcal{D}}_n\Bigg] &= \int \int \Pr\left\{T^{\textsc{prg}*}_{i'} > \tilde{t}^{(b)} \mid T^{\textsc{prg}*}_{i'} > t^{(b)}, \boldsymbol{u}_{i'}, \boldsymbol{\theta}\right\} \\
    &\qquad\qquad \Pr\left[\boldsymbol{u}_{i'} \mid T^{\textsc{prg}*}_{i'} > t^{(b)}, \boldsymbol{\mathcal{Y}}_{i'}\left\{t^{(y)}\right\}, \boldsymbol{\theta}\right] \\
    &\qquad\qquad \Pr(\boldsymbol{\theta \mid \boldsymbol{\mathcal{D}}_n}) d\boldsymbol{u}_{i'} d\boldsymbol{\theta}.
\end{align*}

By conditioning on the most recent conducted biopsy before the current visit, $t^{(b)}$, instead of the most recent tentative biopsy $\tilde{t}^{(b)}$, the biopsy schedule for a patient can be generated efficiently in the sense that the risk estimates at all future clinical visits can be derived from the same risk curve. 

As, over time, additional biomarker information becomes available, the posterior distribution of the random effects must be updated since it depends on $\boldsymbol{\mathcal{Y}}_{i'}\left\{t^{(y)}\right\}$ (where $t^{(y)}$ is now the time of the new biomarker measurements). Consequently, the risk estimates and proposed biopsy schedule are dynamically updated throughout a patient's follow-up.

In the above, we base the decision to schedule a biopsy on the risk of progression but not on the risk of the competing event, since biopsies are not needed to ``detect" early treatment initiation. In other applications, however, the occurrence of multiple competing events may be determined by the same diagnostic test. In that case, the cumulative risk for all of these events should be considered when determining whether the test should be performed or deferred.

    

\begin{figure}[H]
    \centering
    \includegraphics[width = 0.49\textwidth]{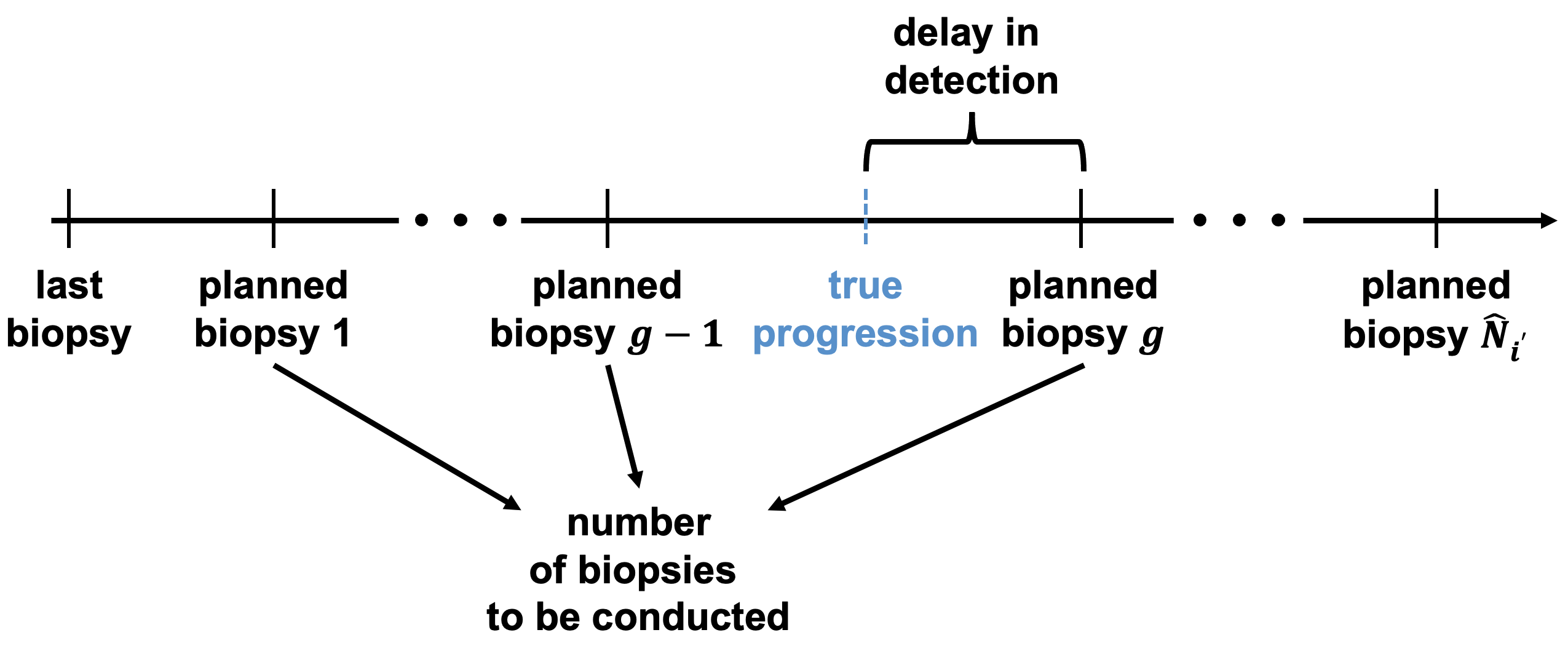}
    \caption{Definition for the number of biopsies ($\textit{Nb}$) and detection delay ($\textit{Dd}$).}
    \label{fig:exampleid}
\end{figure}

The choice of the risk threshold influences the resulting biopsy schedule. A higher risk threshold results in fewer biopsies but a higher risk of missing the optimal treatment window. Since patients progress at different rates, different risk thresholds may be appropriate. We follow the idea of Tomer et al.\cite{Ani2022} to choose a suitable patient-specific risk threshold by balancing the expected number of biopsies and expected delay in detection (see Figure~\ref{fig:exampleid}) and adapt their proposed methodology to the setting with competing risks. 

Given the time of the most recent biopsy $t^{(b)}$ (e.g., $t^{(b)}=0$, if we are at baseline and generate the schedule for the first time), the horizon time $t_V$ and a proposed biopsy schedule $\boldsymbol{\mathcal{B}}_{i'}^{\phi}=\{\tilde{t}^{(b)}_1, \dots, \tilde{t}^{(b)}_{\hat{N}_{i'}}\}$ containing $\hat{N}_{i'}$ planned biopsies based on a threshold $\phi$, the number of biopsies ($\textit{Nb}$) for a patient $i'$ who experiences progression or early treatment is
\begin{align*}
    \textit{Nb}(\boldsymbol{\mathcal{B}}^{\phi}_{i'}) = g, \ &\text{if } \tilde{t}^{(b)}_{g-1} < T^{\textsc{prg}*}_{i'} \leq \tilde{t}^{(b)}_{g},  T^{\textsc{prg}*}_{i'} < T^{\textsc{trt}*}_{i'}\ \text{or} \\
    \textit{Nb}(\boldsymbol{\mathcal{B}}^{\phi}_{i'}) = g-1, \ &\text{if } \tilde{t}^{(b)}_{g-1} < T^{\textsc{trt}*}_{i'} \leq \tilde{t}^{(b)}_{g}, T^{\textsc{trt}*}_{i'}<T^{\textsc{prg}*}_{i'}.
\end{align*}
Patients who are detected with cancer progression before $t_V$ within $\hat{N}_{i'}$ scheduled biopsies, the detection delay ($\textit{Dd}$) is expressed as
\begin{align*}
    \textit{Dd}(\boldsymbol{\mathcal{B}}^{\phi}_{i'}) = \tilde{t}^{(b)}_{g} - T^*_{i'}, \ \text{if } \tilde{t}^{(b)}_{g-1} < T^{\textsc{prg}*}_{i'} \leq \tilde{t}^{(b)}_{g}, T^{\textsc{prg}*}_{i'} < T^{\textsc{trt}*}_{i'},
\end{align*}
with $g \in \{1, \dots, \hat{N}_{i'}\}$. For patients who are censored or start early treatment, the detection delay is undefined as cancer progression does not occur. Because our interest is in the detection of progression and early treatment is not the focus in evaluating a biopsy schedule, this is not regarded as a restriction in our setting. 

Since the actual number of biopsies to be conducted is unknown when creating a schedule and the true progression time is generally unknown in practice, we use the expected number of biopsies and expected delay in detection instead. The expected number of biopsies is:
\begin{align*}
    E\{\textit{Nb}(\boldsymbol{\mathcal{B}}^{\phi}_{i'})\} = \sum^{\hat{N}_{i'}}_{g = 1} g \times \Pr\{\tilde{t}^{(b)}_{g-1} < T^{\textsc{prg}*}_{i'} \leq \tilde{t}^{(b)}_{g} \mid T^{\textsc{prg}*}_{i'} \leq \tilde{t}^{(b)}_{\hat{N}_{i'}}\},
\end{align*}
where
\begin{align*}
    \Pr\{\tilde{t}^{(b)}_{g-1} < T^{\textsc{prg}*}_{i'} \leq \tilde{t}^{(b)}_{g} \mid T^*_{i'} \leq \tilde{t}^{(b)}_{\hat{N}_{i'}}\} = \quad\frac{\Pi^{(\textsc{prg})}_{i'}\left\{\tilde{t}^{(b)}_{g} \mid t^{(b)}, t^{(y)}\right\} - \Pi^{(\textsc{prg})}_{i'}\left\{\tilde{t}^{(b)}_{g-1} \mid t^{(b)}, t^{(y)}\right\}} {\Pi^{(\textsc{org})}_{i'}\left\{\tilde{t}^{(b)}_{\hat{N}_{i'}} \mid t^{(b)}, t^{(y)}\right\}}\ ,
\end{align*}
and the expected detection delay can be calculated as:
\begin{align*}
    E\{\textit{Dd}(\boldsymbol{\mathcal{B}}^{\phi}_{i'})\} = \sum^{N_{i'}}_{g = 1} \Big[\left\{\tilde{t}^{(b)}_{g} - E(T^{\textsc{prg}*}_{i'} \mid \tilde{t}^{(b)}_{g-1}, \tilde{t}^{(b)}_{g}, t^{(y)})\right\} \times \Pr\{\tilde{t}^{(b)}_{g-1} < T^{\textsc{prg}*}_{i'}\leq \tilde{t}^{(b)}_{g} \mid T^{\textsc{prg}*}_{i'} < \tilde{t}^{(b)}_{\hat{N}_{i'}}\}\Big] ,
\end{align*}
where the expected progression time for subject $i'$ between two biopsies $\tilde{t}^{(b)}_{g-1}$ and $\tilde{t}^{(b)}_{g}$ is
\begin{align*}
    E(T^{\textsc{prg}*}_{i'} \mid \tilde{t}^{(b)}_{g-1}, \tilde{t}^{(b)}_{g}, t^{(y)}) &=  \tilde{t}^{(b)}_{g-1} + \int^{\tilde{t}^{(b)}_{g}}_{\tilde{t}^{(b)}_{g-1}}\Pr\Bigg[T^{\textsc{prg}*}_{i'} \geq \nu \mid \tilde{t}^{(b)}_{g-1} < T^{\textsc{prg}*}_{i'} \leq \tilde{t}^{(b)}_{g}, \boldsymbol{\mathcal{Y}}_{i'}\left\{\tilde{t}^{(y)}\right\}, \boldsymbol{\mathcal{D}}_n\Bigg]d\nu \\
    &= \tilde{t}^{(b)}_{g-1} + \int^{\tilde{t}^{(b)}_{g}}_{\tilde{t}^{(b)}_{g-1}} \frac{\Pr\left[\nu \leq T^{\textsc{prg}*}_{i'} \leq \tilde{t}^{(b)}_{g} \mid \boldsymbol{\mathcal{Y}}_{i'}\left\{t^{(y)}\right\}, \boldsymbol{\mathcal{D}}_n\right]}{\Pr\left[\tilde{t}^{(b)}_{g-1} < T^{\textsc{prg}*}_{i'} \leq \tilde{t}^{(b)}_{g} \mid \boldsymbol{\mathcal{Y}}_{i'}\left\{t^{(y)}\right\}, \boldsymbol{\mathcal{D}}_n\right]}d\nu.
\end{align*}
Since the biopsy scheduling processes only serve for the patients who adhere the planned clinical visits (i.e., never start early treatment), it is assured that their true time of early treatment is always later than that of cancer progression. As proved in (\ref{eq:risk_notrt}), the true event is always cancer progression (i.e., $T^*_i = T^{\textsc{prg}*}_{i'}<T^{\textsc{trt}*}_{i'}$), which is omitted in all the above-mentioned formulas.

\begin{figure}[H]
    \centering
    \includegraphics[width = 0.49\textwidth]{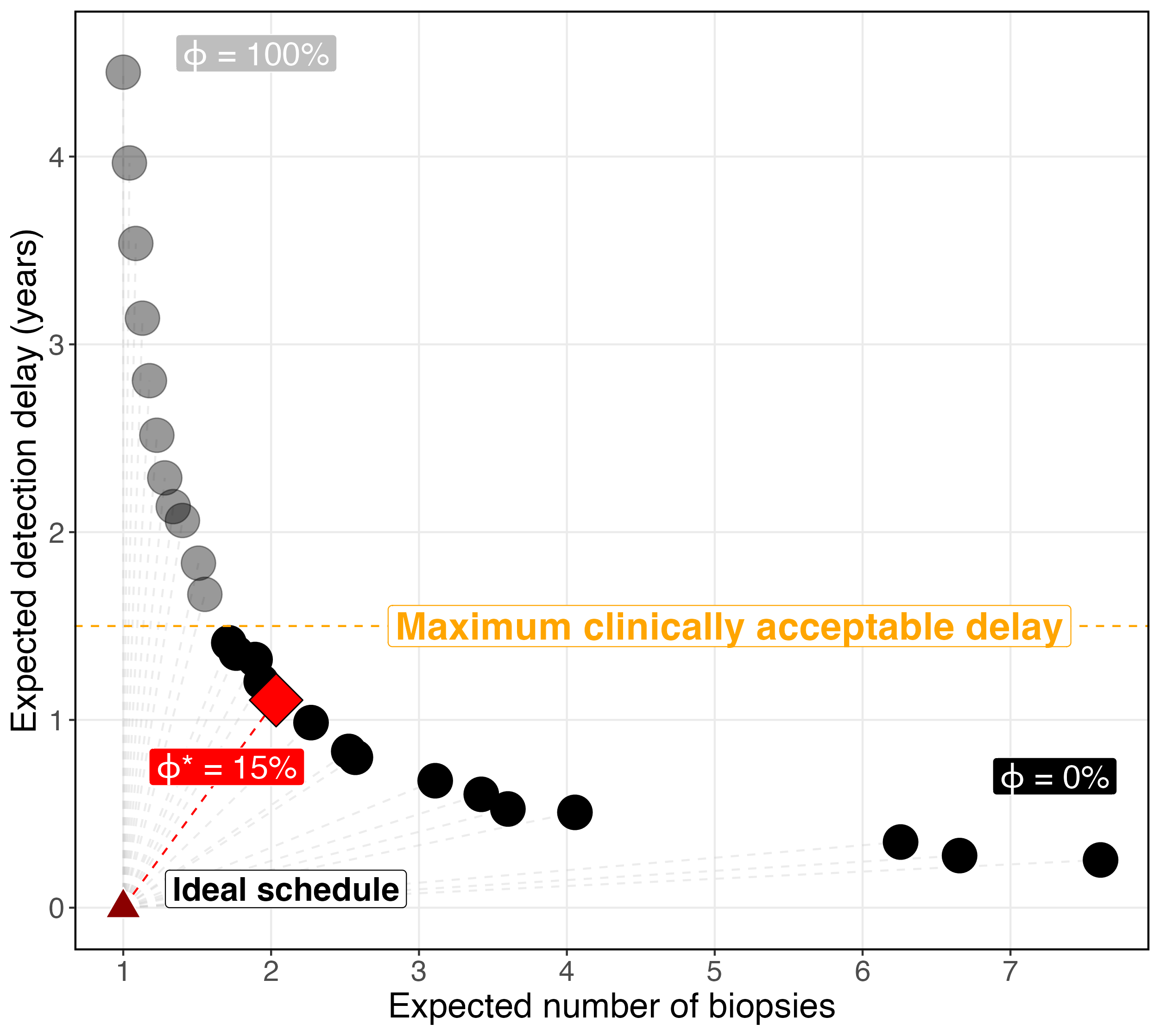}
    \caption{Choice of the optimal risk threshold based on the expected number of biopsies and detection delay.}
    \label{fig:exampleloss}
\end{figure}

For determining the patient-specific risk threshold, we utilize a grid search, creating schedules with a range of thresholds, each with its own expected number of biopsies and delay. Which of the resulting schedules is optimal depends on how the number of biopsies are weighted against the delay. Using weights ($\lambda_1$ and $\lambda_2$) we can then define the optimal risk threshold $\phi$ for patient $i'$ as the one that results in a schedule that minimizes the Euclidean distance of the expected number of biopsies and delay to the ideal schedule in which a single biopsy is performed at the time of cancer progression. This is visualized in Figure~\ref{fig:exampleloss} and can be determined as

\begin{align*}
    \phi_{i'}^*\left\{\tilde{t}^{(b)}\right\} = \argmin_{\phi_{i'} \in [0,1]} \sqrt{\lambda_1 \times \left[E\{\textit{Nb}(\boldsymbol{\mathcal{B}}^{\phi}_{i'})\} - 1\right]^2 + \lambda_2 \times \left[ E\{\textit{Dd}(\boldsymbol{\mathcal{B}}^{\phi}_{i'})\}\right]^2}.
\end{align*}
For simplicity, we fix the weights $\lambda_1$ and $\lambda_2$ to one. In practice, ensuring the detection delay does not exceed a particular value may be desirable. This can be added as a constraint to the optimization. As over time, new biomarker measurements are collected, the optimization process is repeated dynamically to provide personal schedules that take into account all available information.

\section{Analysis of the Canary PASS Data} \label{sec:analysis}

We analyzed the Canary PASS data using two ICJMs. The primary model (ICJM 1) was based on the PSA as the main predictor of cancer progression. PSA trajectories were modeled with a Student's-t mixed model, indicated in (\ref{eq:psa}). Besides, the estimated (underlying) value of PSA at the time of an event, the time-to-event component included the change in the estimated PSA over the previous year, and the patient's age and PSA density at baseline.

To explore the role of the core ratio, we extended ICJM 1 with a binomial mixed model for the core ratio, indicated in (\ref{eq:cr}), and included the estimated value of this ratio in the time-to-event component (ICJM 2).

The most relevant parameter estimates from both models are summarized in Table \ref{Tab:modelresult}. For the full result table, see Web Appendix 4.3. The results from the survival component of the primary model reveal that with a one unit increase in $\log(\text{baseline PSA density})$, the estimated risk of progression increased by a factor 1.66 (95\% credible interval, CI: 1.25-2.16). An increase of the $\log_2(\text{PSA} + 1)$ by one had a limited effect on the progression-specific risk (HR=1.14, 95\% CI: 0.89-1.43) while increasing the change in $\log_2(\text{PSA} + 1)$ over the previous year by one led to a 18.52-fold increase in the risk. In ICJM2, the hazard ratio (HR) for $\log(\text{baseline PSA density})$ was reduced to 1.35 (95\% CI: 0.97-1.79) while the impact of the $\log_2(\text{PSA} + 1)$ value was amplified (HR increased to 1.30, 95\% CI: 1.02-1.65) and that from its yearly change was weakened (HR decreased to 5.21, 95\% CI: 1.35-19.78). Moreover, a one-unit increase in the $\text{logit[E(core ratio)]}$ led to a 3.01 (95\% CI: 2.48-3.62)-fold increment in the progression-specific risk. 

\begin{table*}
    \centering
    \setlength\extrarowheight{3pt}
    \caption{Posterior means and 95\% credible intervals from the joint model for the Canary PASS data.}
    \begin{tabular}{p{6cm}cccc}
        \toprule
        \multirow{2}{*}{\textbf{Parameters}} & \multicolumn{2}{c}{\textbf{ICJM 1 (PSA)}} & \multicolumn{2}{c}{\textbf{ICJM 2 (PSA \& core ratio)}} \\ \cline{2-5}
         & \textbf{HR} & \textbf{95\% CI} & \textbf{HR} & \textbf{95\% CI} \\ \midrule
        \multicolumn{5}{l}{\textbf{Progression-specific survival component}} \\ 
        $\log(\text{PSA density})$ & 1.66 & [1.25, 2.16] & 1.35 & [0.97, 1.79] \\
        $\log_2(\text{PSA} + 1)$ value & 1.14  & [0.89, 1.43] & 1.30 & [1.02, 1.65]\\ 
        $\log_2(\text{PSA} + 1)$ yearly change & 18.52 & [5.73, 60.17] & 5.21 & [1.35, 19.78]\\
        $\text{logit[E(core ratio)] value}$ & - & - & 3.01 & [2.48, 3.62] \\
        \hline
        \multicolumn{5}{l}{\textbf{Treatment-specific survival component}} \\ 
        $\log(\text{PSA density})$ &  1.25 & [0.83, 1.88] & 0.78 & [0.46, 1.30] \\
        $\log_2(\text{PSA} + 1)$ value &  1.52 & [1.09, 2.11] & 1.71 & [1.14, 2.54] \\
        $\log_2(\text{PSA} + 1)$ yearly change &  9.13 & [1.17, 67.12] & 1.67 & [0.18, 16.70] \\ 
        $\text{logit[E(core ratio)] value}$ & - & - & 4.64 & [3.31, 6.61]\\ \bottomrule
        \multicolumn{5}{l}{\small CI: credible interval; HR: hazard ratio; Note: the longitudinal components are not shown.}
    \end{tabular}
    \label{Tab:modelresult}
\end{table*}

Notably, the $\log_2(\text{PSA} + 1)$ and $\text{logit[E(core ratio)]}$ also played important roles in the risk of initiating early treatment. Increasing the $\log_2(\text{PSA} + 1)$ and $\text{logit[E(core ratio)]}$ by one unit resulted in an increased risk of early treatment by a factor of 1.71 (95\% CI: 1.14-2.54) and 4.64 (95\% CI: 3.31-6.61), respectively. This indicates that if a patient has a higher PSA level or a larger core ratio in a biopsy, he does not only have a higher probability of cancer progression, but is also more likely to start early treatment.

\begin{figure*}
   \centering
   \subfigure[]{\includegraphics[width = 0.49\textwidth]{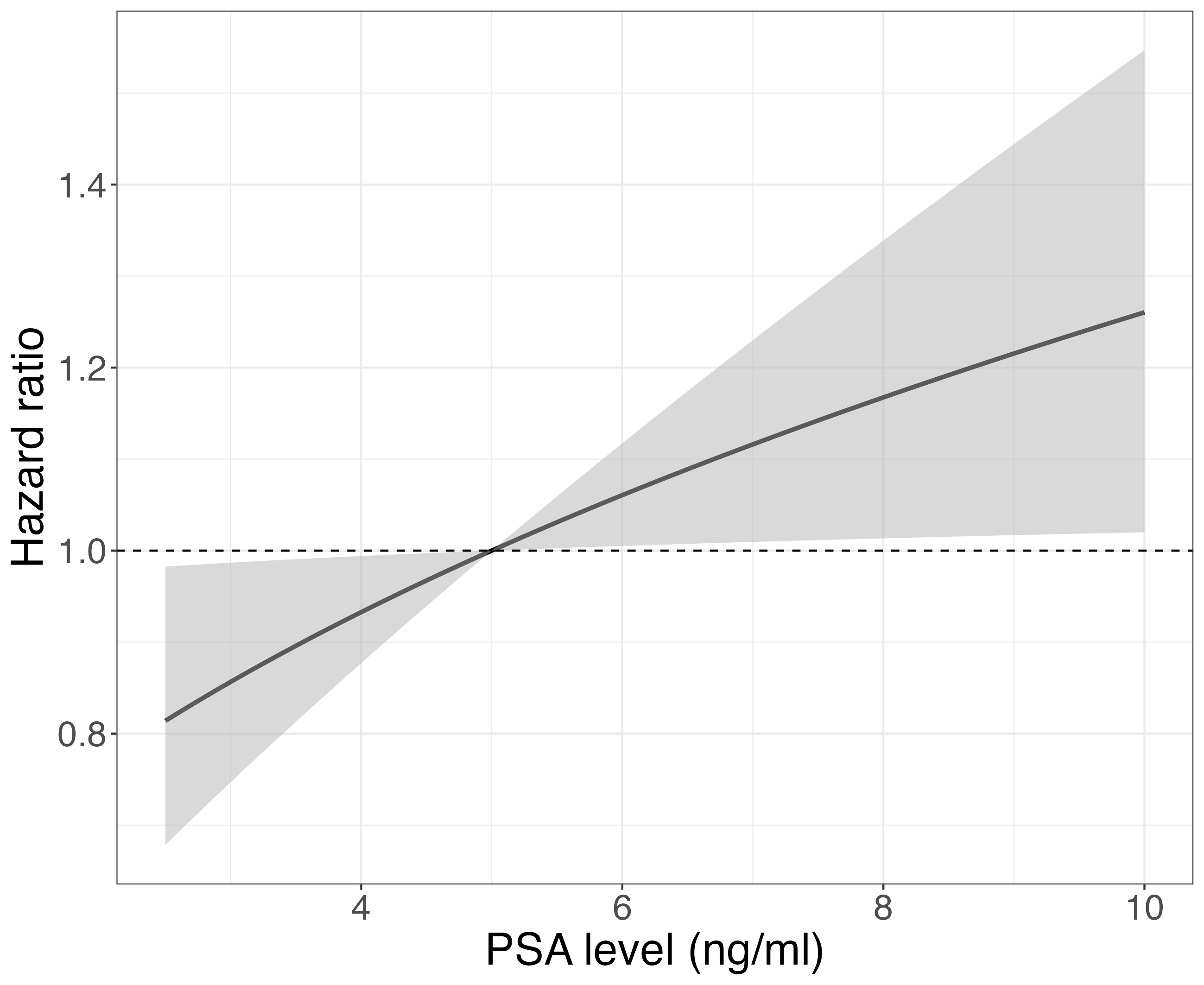}}
   \subfigure[]{\includegraphics[width = 0.49\textwidth]{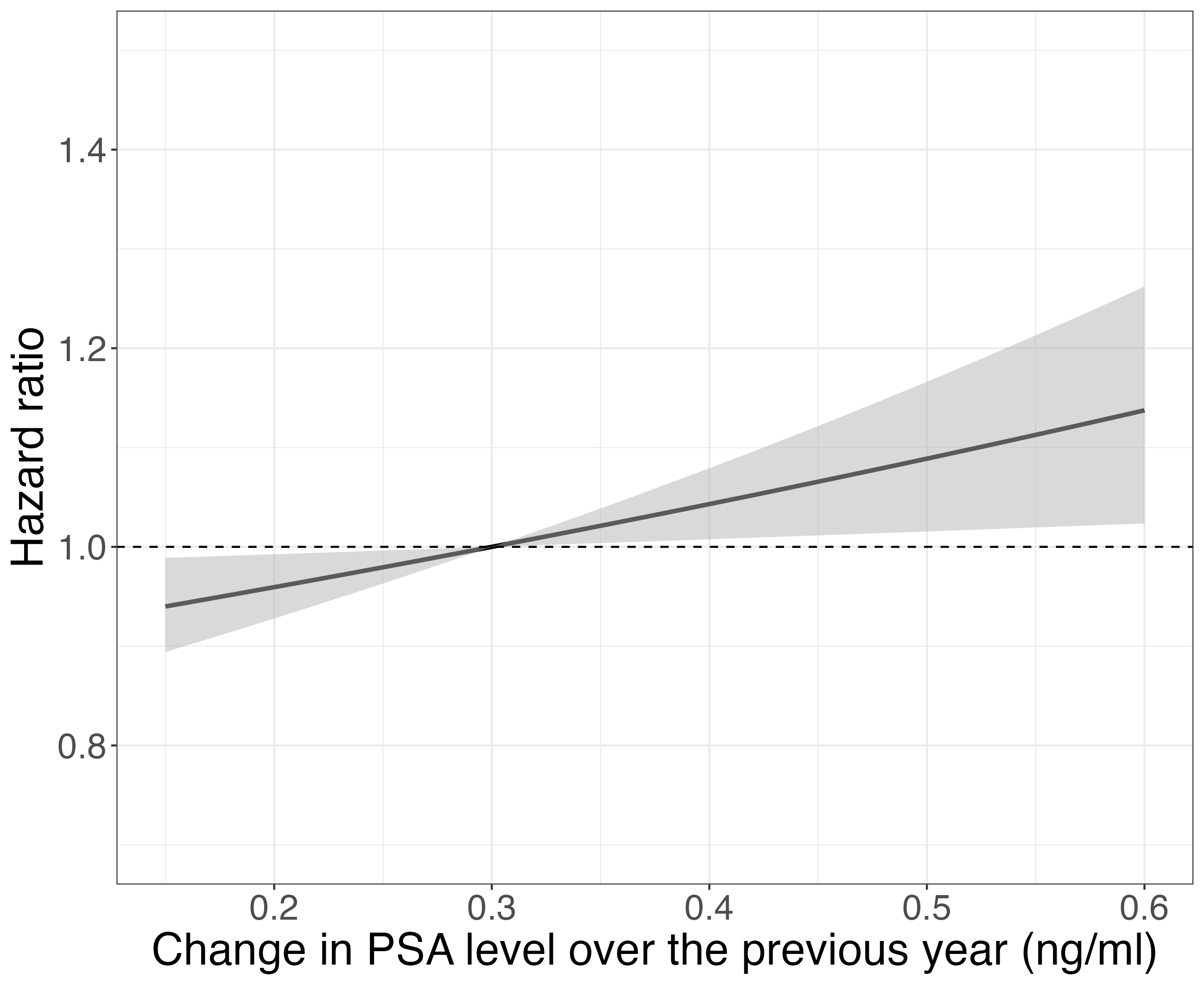}}
   \caption{Effect plot for the impact of (a) the value of PSA and (b) the change in PSA over the previous year on the estimated risk of progression with reference to (a) PSA value 5 ng/ml and (b) change of 0.3 ng/ml, other covariates remaining constant.}
   \label{fig:effplot}
\end{figure*}

\begin{figure}[H]
   \centering
   \includegraphics[width=0.49\textwidth]{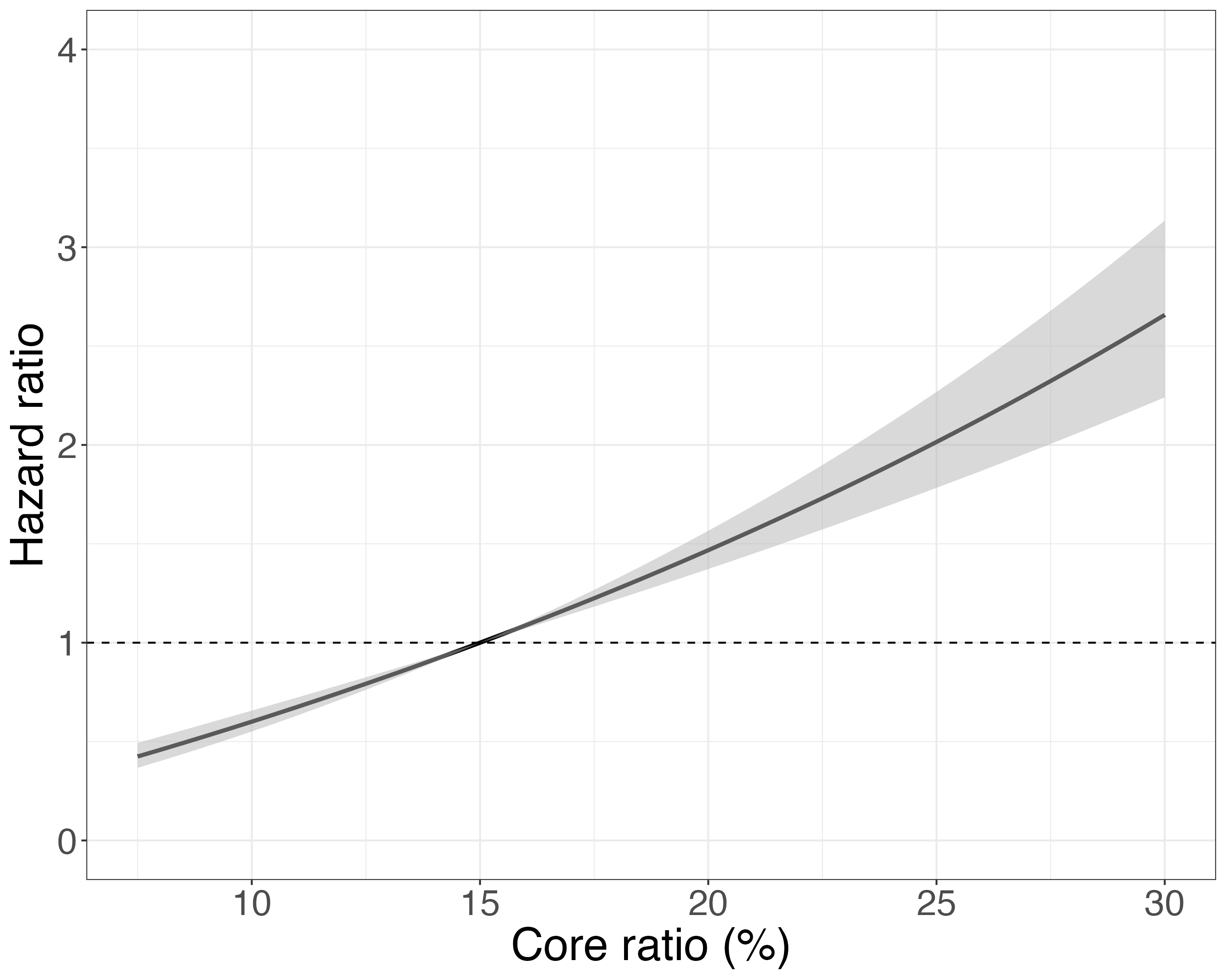}
   \caption{Effect of the core ratio value (contrast to a core ratio of 15\%) on the risk of progression, other covariates remaining constant.}
   \label{fig:effplotcrvalue}
\end{figure}

To facilitate the interpretation of the results in light of the transformed scale, we present effect plots in the original scale (Figure~\ref{fig:effplot} and Figure~\ref{fig:effplotcrvalue}). In the effect plots, the mean PSA level of 5 ng/ml and the mean PSA yearly change of 0.3 ng/ml for patients with median age of 62 were taken as a reference. Figure~\ref{fig:effplot}(a) shows that doubling the PSA level (to 10 ng/ml) resulted in a hazard ratio for cancer progression of 1.26 (95\% CI: 1.02-1.55), while halving the PSA level resulted in a 19\% (95\% CI: 2\%-32\%) decrease of the risk, assuming other covariates remaining constant. As shown in Figure~\ref{fig:effplot}(b), doubling the PSA yearly change (to 0.6 ng/ml) led to a 0.14 (95\% CI: 0.02-0.26)-fold increase of the risk of progression, while halving the PSA yearly change resulted in a 6\% (95\% CI: 1\%-11\%) decrease of the risk. 

The estimated effect of the core ratio is visualized in Figure~\ref{fig:effplotcrvalue}. The observed average core ratio of 15\% was taken as reference. Doubling the core ratio to 30\% raised the progression-specific risk by a factor of 2.66 (95\% CI: 2.24-3.13) while halving the core ratio lowered the risk by 58\% (95\% CI: 51\%-63\%).

The efficacy of the personalized biopsy schedules cannot be directly evaluated on the observed data since true progression time in practice is always unknown. Therefore, a simulation study was performed and is described in the following section.


\section{Simulation} \label{sec:simulation}

We conducted a simulation study to evaluate the efficiency of the proposed personalized biopsy schedules and compare it with that of two existing fixed schedules. The different schedules are evaluated based on the indicators introduced above, the number of biopsies the patients undergo from entering AS and the delay between cancer progression and detection of progression. Since, in this simulation, the progression time is known, the actual delay and number of conducted biopsies are used in the evaluation and comparison (while the expected values are used for generating the schedules). In Section~\ref{ss}, the simulation procedures and scenarios are presented. The results are summarized in Section~\ref{sr}.

\subsection{Simulation Setting} \label{ss}

Data were simulated from the primary ICJM fitted on the Canary PASS data (ICJM 1 in Section~\ref{sec:analysis}). In total, 200 datasets were created, each of which was split into a training set of 300 subjects and a test set of 200 subjects. Each test set contained 100 subjects with progression and 100 subjects without progression. We compared the personalized schedules with a fixed schedule of yearly biopsies and the schedule from the Canary PASS protocol (see Section~\ref{sec:data}). Since the ICJM is mainly used for prediction in practice, one important scenario can be that at the clinical visit when the patient experiences a biopsy, the doctor wants to predict the risk of cancer progression in the future (i.e., $t^{(b)}=t_v$). In such a case, (\ref{eq:riskformula}) can be simplified to 
\begin{equation}
\begin{aligned}[b]
    \frac{\displaystyle{\int}^{t^{(p)}}_{t^{(b)}} h^{(\textsc{prg})}_{i'}(\nu)S_{i'}(\nu)d\nu}{S_{i'}\{t^{(b)}\}}, \ \text{if } t^{(b)}=t_v,
\end{aligned}
\label{eq:form_riskeval}
\end{equation}
where $S_{i'}(\cdot)$ is the overall survival at a certain time point. We also evaluated the predictive performance of the ICJM based on the above-mentioned formula, see Web Appendix 3.2.

For each simulated dataset, we fitted the ICJM on the training set and generated personalized schedules for the subjects in the test set. All subjects in the test set were assumed to have an initial (progression-negative) biopsy when entering AS (i.e., at time zero), visit the clinics every six months, and have PSA levels evaluated every three months. Schedules were generated for a horizon time of 10 years. The maximum acceptable expected delay for the personalized schedules was fixed to 1.5 years. 

\subsection{Simulation Results} \label{sr}

\begin{figure*}
    \centering
    \includegraphics[width = 0.99\textwidth]{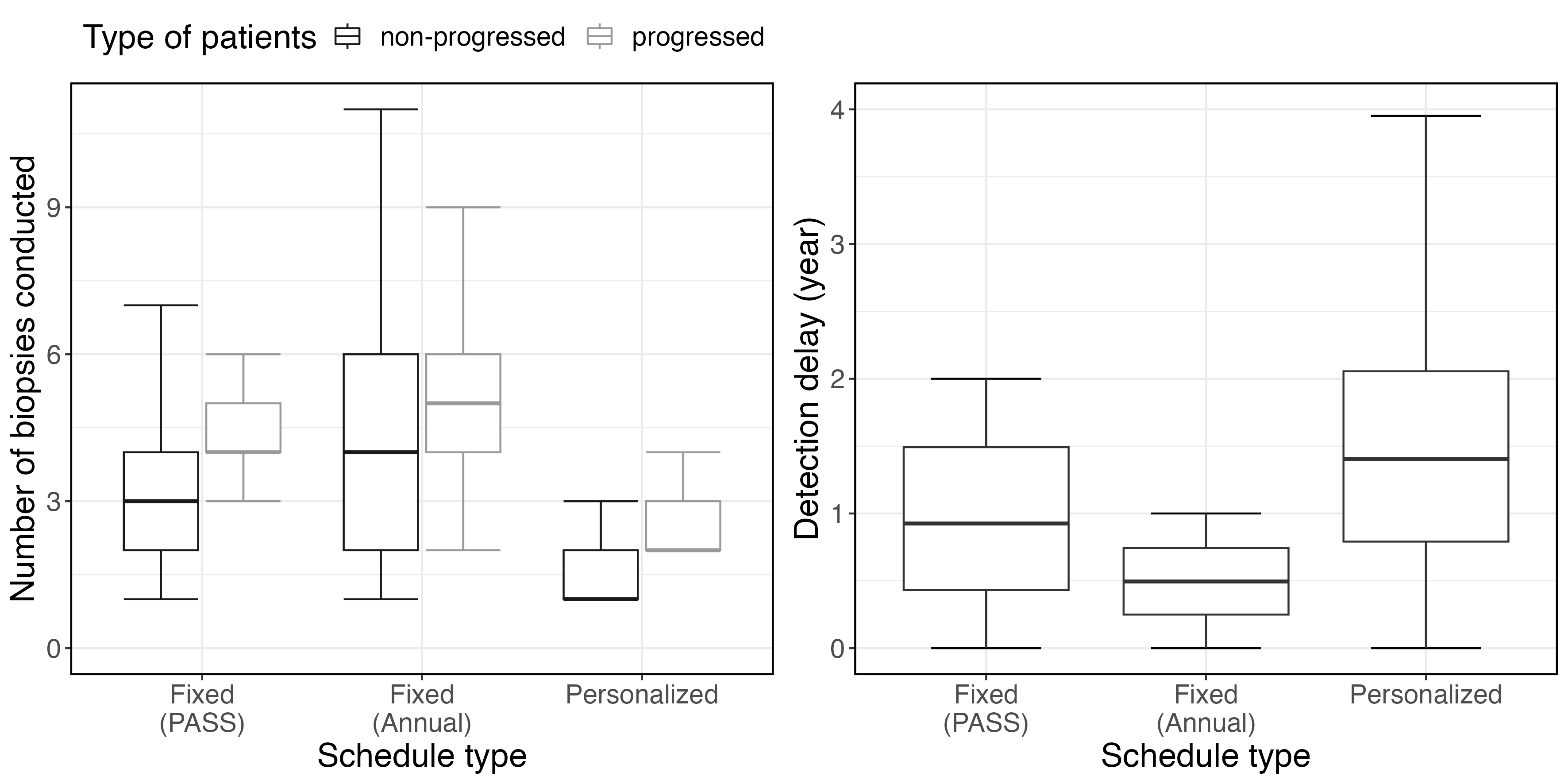}
    \caption{Number of biopsies (including the biopsy conducted when entering the AS) for all $200 \times 200$ patients and detection delay for $200 \times 100$ progressed patients in the test sets . }
    \label{fig:simres}
\end{figure*}

The results of our simulation study are visualized in Figure~\ref{fig:simres}. For patients whose cancer progression had been detected, the median number of biopsies (including the initial biopsy) was four in the PASS schedule and five in the annual schedule. The personalized biopsy schedules required only a median of two biopsies, resulting in an average decrease of 1.77 (41\%) and 2.57 (48\%) biopsies per patient compared to the PASS and annual fixed schedule, respectively. As a trade-off, the personalized schedules resulted in a moderately longer detection delay of 1.42 years (median) compared to 0.93 years delay for the PASS schedule and 0.50 years delay for the annual biopsies. For patients who did not experience cancer progression before the horizon time, we observed that the personalized schedules reduced the number of biopsies by, on average, 1.55 (44\%) and 2.58 (52\%) per patient compared to the PASS biopsy schedule and the annual biopsy schedule, respectively. 

\section{Discussion} \label{sec:discussion}

In this study, we proposed the interval-censored cause-specific joint model to estimate the patient-specific risk of cancer progression in AS of prostate cancer patients, using longitudinal biomarkers while considering the competing risk of early treatment initiation. The patient-specific risk profiles were then used to generate personalized biopsy schedules. In our simulation study, the personalized schedules reduced the number of biopsies per patient on average by 41\%-52\% compared to two fixed schedules. As a trade-off, the personalized schedules lead to slightly longer detection delays.

The ICJM retains the qualities of the standard joint model to handle endogeneity and random measurement error in the longitudinal outcomes but also takes into account interval-censored clinical outcomes and competing risks. Although our model was developed for prostate cancer patients on AS, it has many applications in various clinical fields since interval censoring due to periodic examinations (e.g., regular screening tests in oncology, and periodic tests in infectious diseases) is common and the outcomes of interest are often contaminated by competing events (e.g., death). Specifically in our case, considering informative censoring due to early treatment initiation aids the ICJM in reducing bias in predicting the risk of cancer progression necessary for informed decision-making. The simulation study showing the efficacy of the methodology for personalized biopsy schedules focused on the PSA as the main predictor of cancer progression. The ICJM can be easily extended with additional longitudinal outcomes and handle the issue of different timelines in different longitudinal biomarkers, as demonstrated in ICJM 2 with the core ratio. Considering more longitudinal outcomes may further improve the predictive ability of the ICJM and thus the quality of the personalized schedules. With the increase in the dimensionality of the ICJM, shrinkage methods may become relevant.\cite{Andrinopoulou2016}

The frequency and timing of biopsies in our personalized schedules are based on a trade-off between the resulting expected number of biopsies and expected delay in detecting progression. That these two indicators are on very different scales makes it challenging to decide how they should be weighted relative to each other. More information on how detection delay is related to clinical outcomes, such as mortality, is needed to better tailor this trade-off to the actual risks associated with a particular schedule. It may also be worth exploring whether this relationship differs by patient characteristics like the patient's age or time since the start of active surveillance. Studies with longer follow-up, for instance, where patients are monitored until death, or microsimulation models,\cite{tiago2017} may help to gain the necessary insights into more comprehensive clinical benefits of personalized scheduling by considering factors such as the costs of treatment and quality-adjusted life years (QALYs).

Further adaptations of the methodology are necessary to reflect the actual clinical setting more closely. For example, the current approach does not yet consider that biopsies are imperfect measurements and may give false-negative results. Moreover, other diagnostic tools, such as magnetic resonance imaging (MRI), are utilized in clinical practice to guide the decision to perform a biopsy. Future extensions might, thus, consider the additional uncertainty due to false-negative biopsies and more complex decision-making processes.

In conclusion, the proposed ICJM allows us to obtain patient-specific dynamic risk predictions for interval-censored events in the presence of competing risks that can be translated to personalized biopsy schedules for prostate cancer patients in AS. The resulting schedules can relieve the burden on patients by considerably reducing the number of biopsies while limiting the delay in detecting cancer progression. 

\section*{Acknowledgement}
The research was funded by the National Institutes of Health (the NIH CISNET Prostate Award CA253910). The authors would also like to show our gratitude to the Canary PASS team and all study participants.

\bibliographystyle{SageV}
\bibliography{myref}

\label{lastpage}
\end{document}


\maketitle

\section{Data}

\subsection{PASS Data}

Table~\ref{Tab:t1} summarizes the relevant subset of the Canary PASS data. 

\begin{table*}[!ht]
    \centering
    \caption{Summary table for the Canary PASS Data.}
    \begin{tabular}{ll}
        \toprule
        \textbf{Item}                            & \textbf{Value} \\ \midrule
        Number of subjects  & 833  \\
        Observation time until progression/treatment (years)$^*$    & 4.35 (2.82-6.18)  \\
        Baseline PSA density$^\ddagger$ ($\text{ng}/\text{ml}^2$)$^\dagger$  & 0.12 (0.10)  \\
        Age at start of AS (years)$^*$  & 62 (57-67)   \\
        Total number of PSA measurements & 8262  \\
        Number of PSA measurements per patient$^*$   & 9 (5-14) \\
        PSA level (ng/ml)$^\dagger$  & 5.10 (3.84)  \\
        Number of positive cores per patient$^*$ & 3 (2-4) \\
        core ratio (\%)$^*$ & 8.33 (0.00-16.67)\\
        Number of biopsies per patient$^*$ & 2 (2-3)  \\
        \bottomrule
        \multicolumn{2}{l}{\small $^*$ median is shown followed by the interval between 25\% quantile and 75\% quantile; } \\
        \multicolumn{2}{l}{\small $^\dagger$ mean is shown with standard deviation in the parentheses; } \\
        \multicolumn{2}{l}{\small $^\ddagger$: PSA density equals to PSA level (ng/ml) divided by prostate volume (ml).}
    \end{tabular}
    \label{Tab:t1}
\end{table*}

\subsection{Time-to-event Outcomes}

\begin{figure}[H]
    \centering
    \includegraphics[width = 0.9\textwidth]{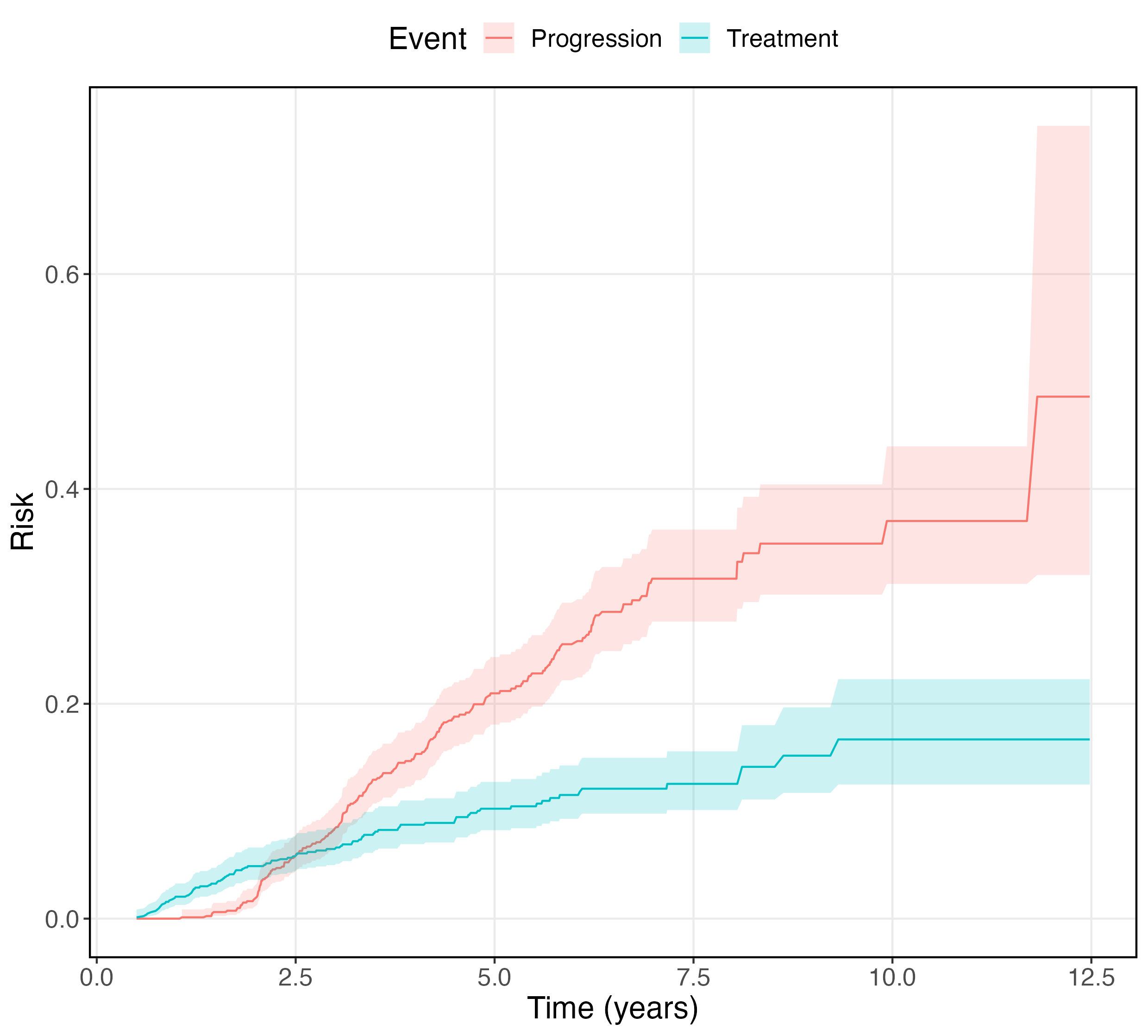}
    \caption{The Aalen-Johansen estimator of progression- and treatment-specific risk in the PASS data.}
    \label{fig:aj}
\end{figure}

The Aalen–Johansen estrimator of the two events' risks in the PASS data is visualized in Figure~\ref{fig:aj}.

\subsection{Longitudinal Outcomes}

\begin{figure}[H]
    \centering
    \subfigure[PSA levels]{\includegraphics[width = 0.9\textwidth]{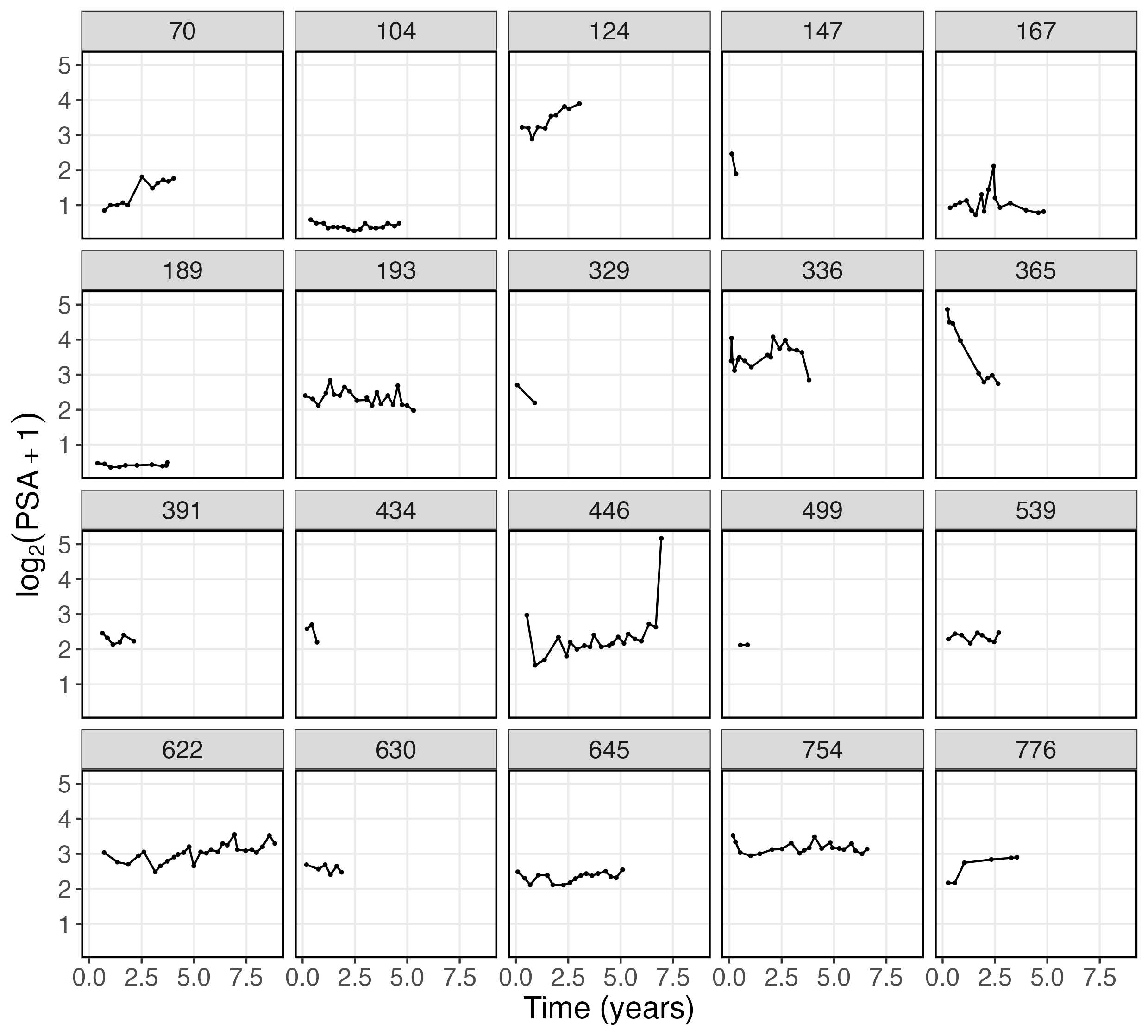}}
    \subfigure[Core ratios]{\includegraphics[width = 0.9\textwidth]{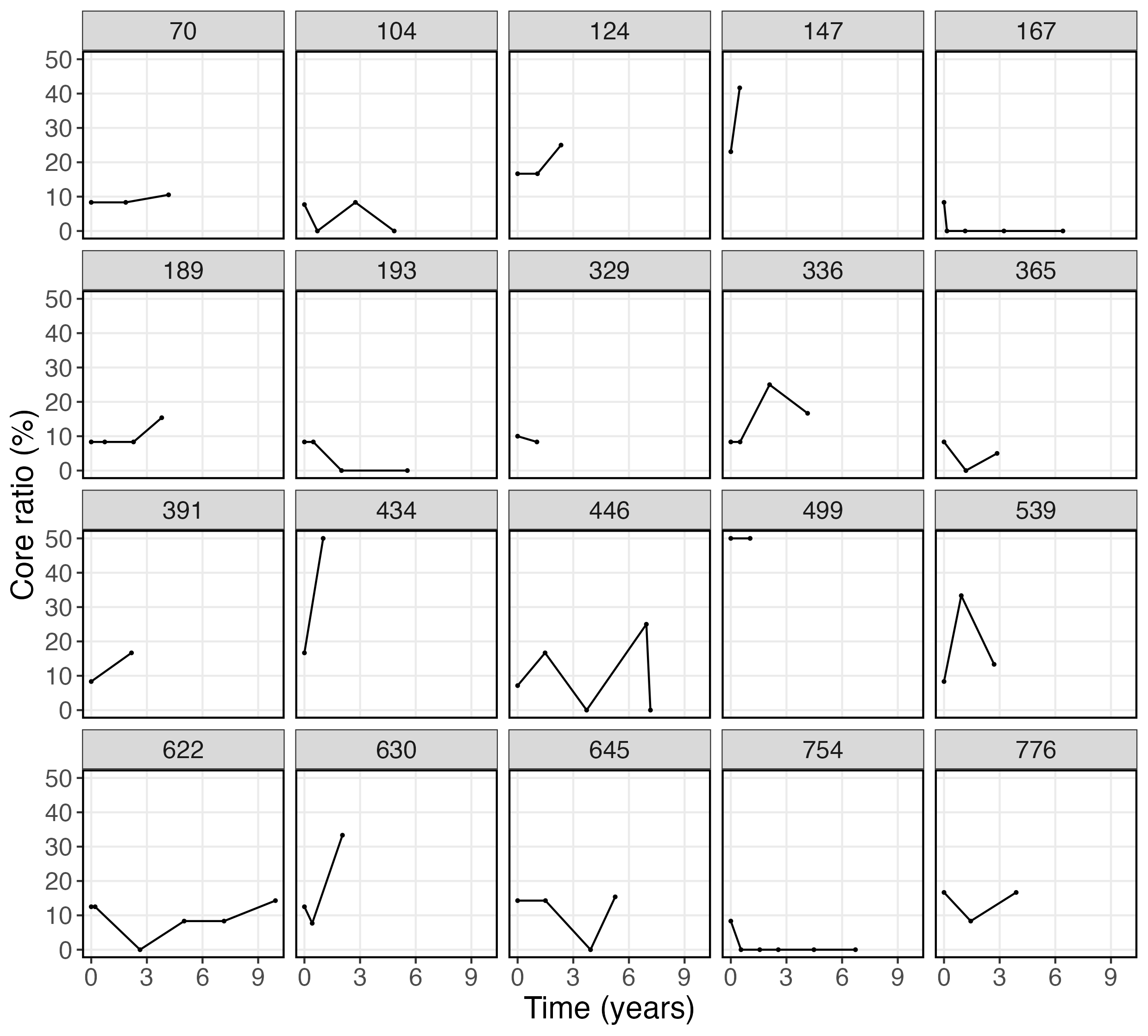}}
    \caption{Observed trajectories of two longitudinal outcomes for 20 randomly selected subjects.}
    \label{fig:observedtrajectory}
\end{figure}

In the Canary PASS data, two longitudinal outcome are available, namely, PSA levels and proportion of cores obtained by a biopsy that contain cancer cells (core ratio). In Figure~\ref{fig:observedtrajectory}, the development of the longitudinal outcomes is displayed for 20 randomly selected patients. The trajectories show non-linear evolutions over time, and vary greatly between patients, which needs to be accommodated in the longitudinal component of the ICJM.

\section{Interval-censored Cause-specific Joint Models (ICJM) for Longitudinal and Time-to-event Data}

In the specification of the ICJM, we use a flexible semi-parametric specification of the baseline hazard using penalized B-splines,
\begin{align*}
    \log h_0^{(k)}(t) = \gamma_{k, h_0, 0} + \sum^A_{a=1}{\gamma_{k, h_0, a}\mathcal{G}_a(t, \boldsymbol{\xi})},
\end{align*}
where $\mathcal{G}_a(t, \boldsymbol{\xi})$ is the $a$-th basis function of a B-splines with knots $\xi_1, \dots, \xi_A$. The number of knots was chosen to be 11. The penalized coefficients for the basis function $\boldsymbol{\gamma}_{k, h_0}$ have the following priors,
\begin{align*}
    p(\boldsymbol{\gamma}_{k, h_0} \mid \tau_{k,h_0}) \propto \tau_{k,h_0}^{\rho(\boldsymbol{M})/2} \exp\left(- \frac{\tau_{k,h_0}}{2}\boldsymbol{\gamma}_{k, h_0}^\top\boldsymbol{M}\boldsymbol{\gamma}_{k, h_0}\right),
\end{align*}
with
\begin{align*}
     \tau_{k,h_0} \sim \text{Gamma}(5, 0.5),
\end{align*}
where $\tau_{k,h_0}$ is the smoothing parameter; $\boldsymbol{M} = \Delta_r^\top \Delta_r + 10^{-6}I$, $\Delta_r$ is the $r$-th difference penalty matrix and $\rho(\boldsymbol{M})$ denotes the rank of $\boldsymbol{M}$. 

\section{Simulation Study}

\subsection{Simulation Setting} \label{subsection:icjm1}

To evaluate the performance of our proposed methodology, we simulated data based on the parameters from the ICJM fitted on the Canary PASS study (ICJM 1, see Section 5). The patients in the training sets are supposed to take biopsies in months 12, 24 and afterwards biennially and PSA measurements every three months, with small variations.

This model had the following structure:
\begin{align*}
    \log_2(\text{PSA}_i + 1)(t) &= m_{\textsc{psa},i}(t) + \epsilon_i(t),\\
                    m_{\textsc{psa},i}(t)  &= \beta_0 + u_{0i} + \sum^3_{p=1}(\beta_p + u_{pi})\mathcal{C}^{(p)}_i(t) + \beta_4(\text{Age}_i - 62), \\
    h_i^{(k)}\left\{t \mid \boldsymbol{\mathcal{M}}_{\textsc{psa},i}(t)\right\} &= h_0^{(k)}(t)\exp\Big[\gamma_k\text{density}_i + f\left\{\boldsymbol{\mathcal{M}}_{\textsc{psa},i}(t), \boldsymbol{\alpha}_{k}\right\}\Big],
\end{align*}
where $\mathcal{C}(t)$is the design matrix for the natural cubic splines (with three degrees of freedom) for time $t$; $\text{Age}_i$ and $\text{density}_i$ refer to the patient's age and PSA density at the start of active surveillance, respectively. Baseline Age was centered by subtracting the median age (62 years) for computational reasons. Both the expected value of PSA and the change in expected PSA over the previous year (where extrapolation was conducted for time points earlier than year one) were included as covariates in the time-to-event component, i.e.,
\begin{align*}
    f\left\{\boldsymbol{\mathcal{M}}_{\textsc{psa},i}(t), \boldsymbol{\alpha}_{k}\right\} = &\  \alpha_{1k,\textsc{psa}}m_{\textsc{psa},i}(t) + \alpha_{2k,\textsc{psa}}\Big\{m_{\textsc{psa},i}(t) - m_{\textsc{psa},i}(t-1)\Big\}.
\end{align*}

The residuals of the longitudinal component were assumed to follow a Student's t distribution with three degrees of freedom \cite{Ani2022},
\begin{align*}
    \epsilon_i(t) \sim t(\frac{1}{\tau_\epsilon}, 3),
\end{align*}
with
\begin{align*}
    \tau_\epsilon \sim \text{Gamma}(0.01, 0.01).
\end{align*}
The prior distributions for the regression coefficients were specified as vague normal distributions,
\begin{align*}
    \beta &\sim \mathcal{N}(0, 100), \\
    \gamma_k &\sim \mathcal{N}(0, 100), \\
    \alpha_{1k,\textsc{psa}}, \alpha_{2k,\textsc{psa}} &\sim \mathcal{N}(0, 100),
\end{align*}
and the variance-covariance matrix of the random effects, $\boldsymbol{\Omega}$, to follow an inverse-Wishart distribution,
\begin{align*}
    \boldsymbol{\Omega} \sim \mathcal{IW}(n_u + 1, \frac{4}{\tau_u}),
\end{align*}
with
\begin{align*}
    \tau_u \sim \text{Gamma}(0.5, 0.01),
\end{align*}
where $n_u$ is the number of coefficients for random effects.

The model was implemented in JAGS \cite{JAGS} and run for 10000 iterations, using a thinning interval of 10, in each of three MCMC chains.

The resulting posterior means used for simulation were
\begin{align*}
    \boldsymbol{\beta} &= [2.34, 0.28, 0.61, 0.95, 0.02]^\top, \\
    \boldsymbol{\Omega} &= \begin{bmatrix}
        0.48 & -0.04 & -0.07 & 0.02 \\
        -0.04 & 0.77 & 0.46 & -0.04 \\
        -0.07 & 0.46 & 1.37 & 1.36 \\
        0.02 & -0.04 & 1.36 & 2.54
    \end{bmatrix}, \\
    \tau_\epsilon &= 47.40, \\
    \boldsymbol{\gamma}_{h_0} &= \begin{bmatrix}
        -6.78 & -5.76 \\
        -4.72 & -4.99 \\
        -2.84 & -4.43 \\
        -1.65 & -4.26 \\
        -1.54 & -4.36 \\
        -1.79 & -4.47 \\
        -1.85 & -4.60 \\
        -1.75 & -4.69 \\
        -1.85 & -4.78 \\
        -2.04 & -4.92 \\
        -2.18 & -5.08 \\
        -2.32 & -5.21
    \end{bmatrix},\\
    \boldsymbol{\gamma} &= [0.50, 0.23], \\
    \boldsymbol{\alpha} &= \begin{bmatrix} 
        0.13 & 0.42 \\
        3.01 & 2.62
    \end{bmatrix}.
\end{align*}

The resulting simulated data matched the observed data well with regard to the rates of cancer progression, early treatment initiation and censoring (Table~\ref{Tab:summarytraining}).

\begin{table}[H]
    \centering
    \caption{Summary of event proportions in the simulated training datasets compared to the observed data.}
    \begin{tabular}{lcc}
        \toprule
        \textbf{Events}  & \textbf{Simulated data}$^\dagger$ (\%) & \textbf{Observed data} (\%) \\ \midrule
        Cancer progression & 28.29 & 21.97 \\
        Treatment & 7.92 & 10.44 \\
        Censoring & 63.79 & 67.59 \\
        \bottomrule
        \multicolumn{3}{l}{$^\dagger$: the average proportions overall training sets are presented.}
    \end{tabular}
    \label{Tab:summarytraining}
\end{table}

\subsection{Evaluation of the ICJM}

Since the quality of the personalized schedules relies on good predictive accuracy, we investigated the prediction error of the ICJM.

We predicted the 2-year cancer progression risk at different starting points (baseline, year 1, 2, 3, 4 and 6) assuming patients just experienced the biopsies, based on (8) from the manuscript, for all patients in the test sets based on the ICJMs fitted on the corresponding training sets and calculated the prediction error (i.e., the difference between the true and predicted risk) in R \cite{R}.

\begin{figure}[H]
    \centering
    \includegraphics[width=0.8\textwidth]{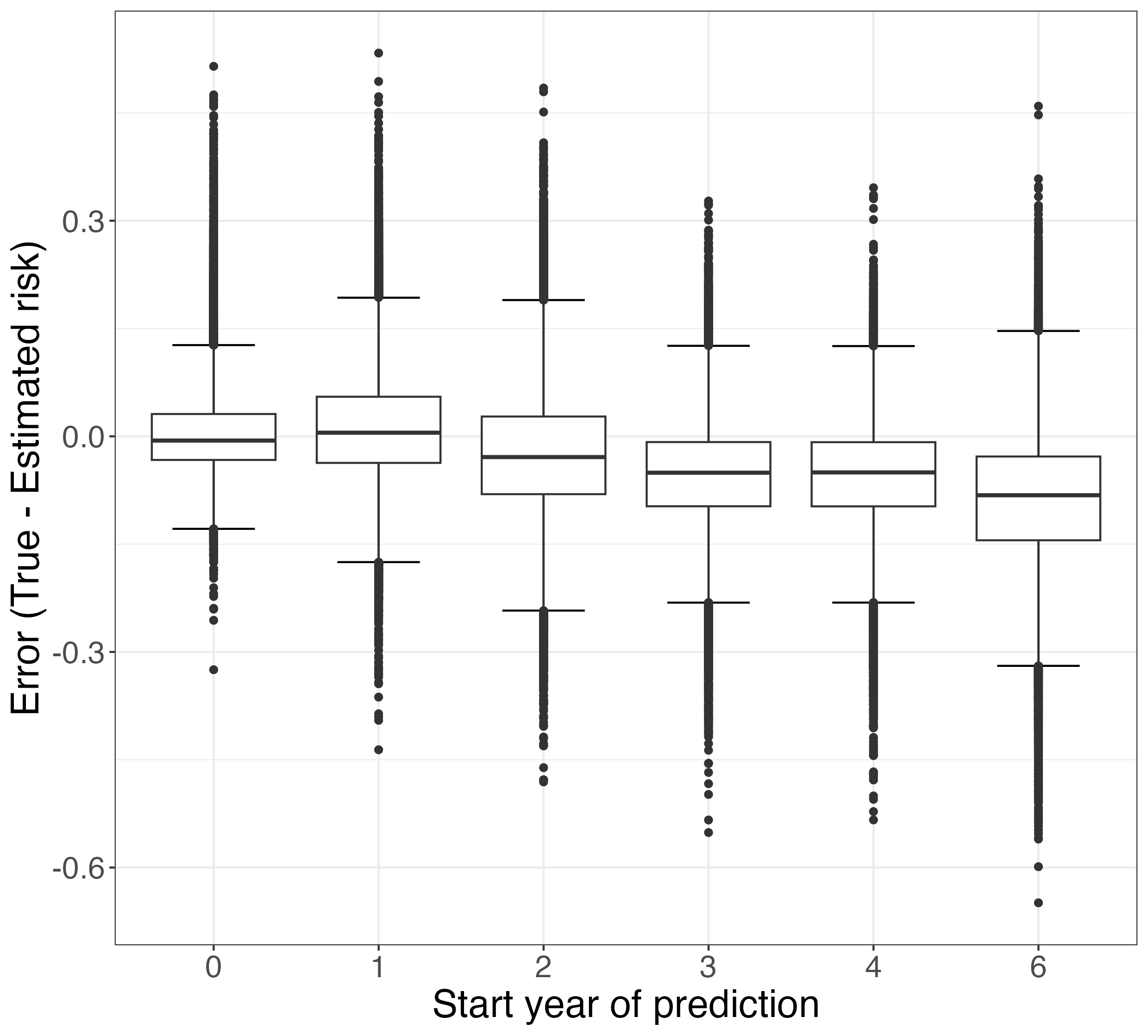}
    \caption{Two-year risk prediction performance evaluation for 200 test sets.}
    \label{fig:evaluation}
\end{figure}

\begin{figure}[H]
    \centering
    \includegraphics[width=0.8\textwidth]{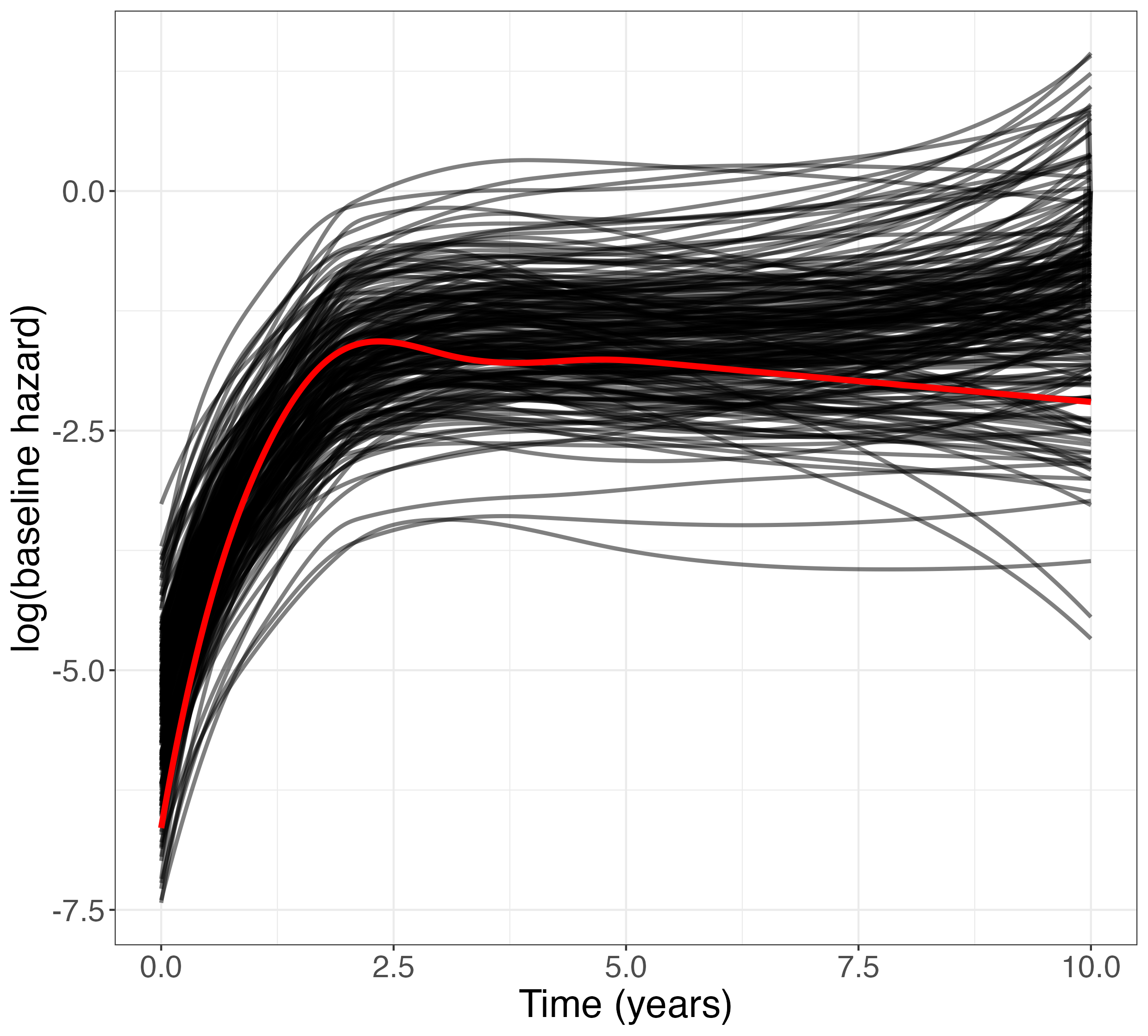}
    \caption{Visualization of the baseline hazards over time (red curve: true baseline hazard ; black curves: estimated baseline hazards).}
    \label{fig:bhazard}
\end{figure}

Figure~\ref{fig:evaluation} visualizes these results; it shows that the ICJM slightly overestimates the 2-year progression risk for later years. Further investigation revealed that this likely stems from the overestimation of the baseline hazard, as shown in Figure~\ref{fig:bhazard}.
The placement of the knots in the spline specification of the baseline hazard is based on the quantiles of the event times in the training data. Since most events occur between years 3 and 4, there is little information to guide the spline fit in the period after year 5.

\section{ICJM for Data Analysis}

The model specification of our primary model, ICJM 1, is detailed in \ref{subsection:icjm1}.

\subsection{Model Specification of ICJM 2} \label{subsection:icjm2}

To investigate the role of the core ratio as a potential predictor for cancer progression and to demonstrate how to incorporate additional biomarkers in the ICJM, we extended ICJM with a binomial mixed model for the core ratio. In this model, we assumed a quadratic evolution over time. The random effects were modelled jointly with the random effects in the model for PSA, i.e.,
\begin{align*}
    \log_2(\text{PSA}_i + 1)(t) &= m_{\textsc{psa},i}(t) + \epsilon_i(t),\\
                    m_{\textsc{psa},i}(t) &= \beta_0 + u_{0i} + \sum^3_p(\beta_p + u_{pi})\mathcal{C}^{(p)}_i(t) + \beta_4(\text{Age}_i - 62), \\
    \text{logit}\left[E\{\text{core ratio}_i(t)\}\right] &= m_{2i}(t), \\
    m_{2i}(t) &= \beta_5 + u_{4i} + (\beta_6 + u_{5i})t + (\beta_7 + u_{6i})t^2,
\end{align*}
where the random effects from the two longitudinal outcomes $\boldsymbol{u}_i = (u_{1i}, \dots, u_{6i})^\top$ are modeled jointly using a multivariate normal distribution, $\boldsymbol{u}_i \sim \mathcal{N}(0, \boldsymbol{\Omega})$.

The survival component of the ICJM was extended to also include the estimated trajectory of the core-ratio, $\boldsymbol{\mathcal{M}}_{\textsc{cr},i}(t)$,
\begin{align*}
                h_i^{(k)}\left\{t \mid \boldsymbol{\mathcal{M}}_{\textsc{psa},i}(t), \boldsymbol{\mathcal{M}}_{\textsc{cr},i}(t)\right\} = h_0^{(k)}(t)\exp\Big[\gamma_k\text{density}_i + f\left\{\boldsymbol{\mathcal{M}}_{\textsc{psa},i}(t), \boldsymbol{\mathcal{M}}_{\textsc{cr},i}(t), \boldsymbol{\alpha}_{k}\right\}\Big],
            \end{align*}
where $\boldsymbol{\alpha}_k = [\alpha_{1k,\textsc{psa}}, \alpha_{2k,\textsc{psa}}, \alpha_{1k,\textsc{cr}}]$ and $f\left\{\boldsymbol{\mathcal{M}}_{\textsc{psa},i}(t), \boldsymbol{\mathcal{M}}_{\textsc{cr},i}(t), \boldsymbol{\alpha}_{k}\right\}$ now also included the expected value of the core ratio,
\begin{align*}
    f\left\{\boldsymbol{\mathcal{M}}_{\textsc{psa},i}(t), \boldsymbol{\mathcal{M}}_{\textsc{cr},i}(t), \boldsymbol{\alpha}_{k}\right\} = &\  \alpha_{1k,\textsc{psa}}m_{\textsc{psa},i}(t) + \alpha_{2k,\textsc{psa}}\Big\{m_{\textsc{psa},i}(t) - m_{\textsc{psa},i}(t-1)\Big\} \\
    &+ \alpha_{1k,\textsc{cr}}m_{\textsc{cr},i}(t).
\end{align*}
ICJM 2 was fitted in JAGS, using 10000 iterations, using a thinning interval of 10, in each of three MCMC chains.

\subsection{Goodness of fit}

\begin{figure}[H]
    \centering
    \subfigure[PSA levels]{\includegraphics[width = 0.8\textwidth]{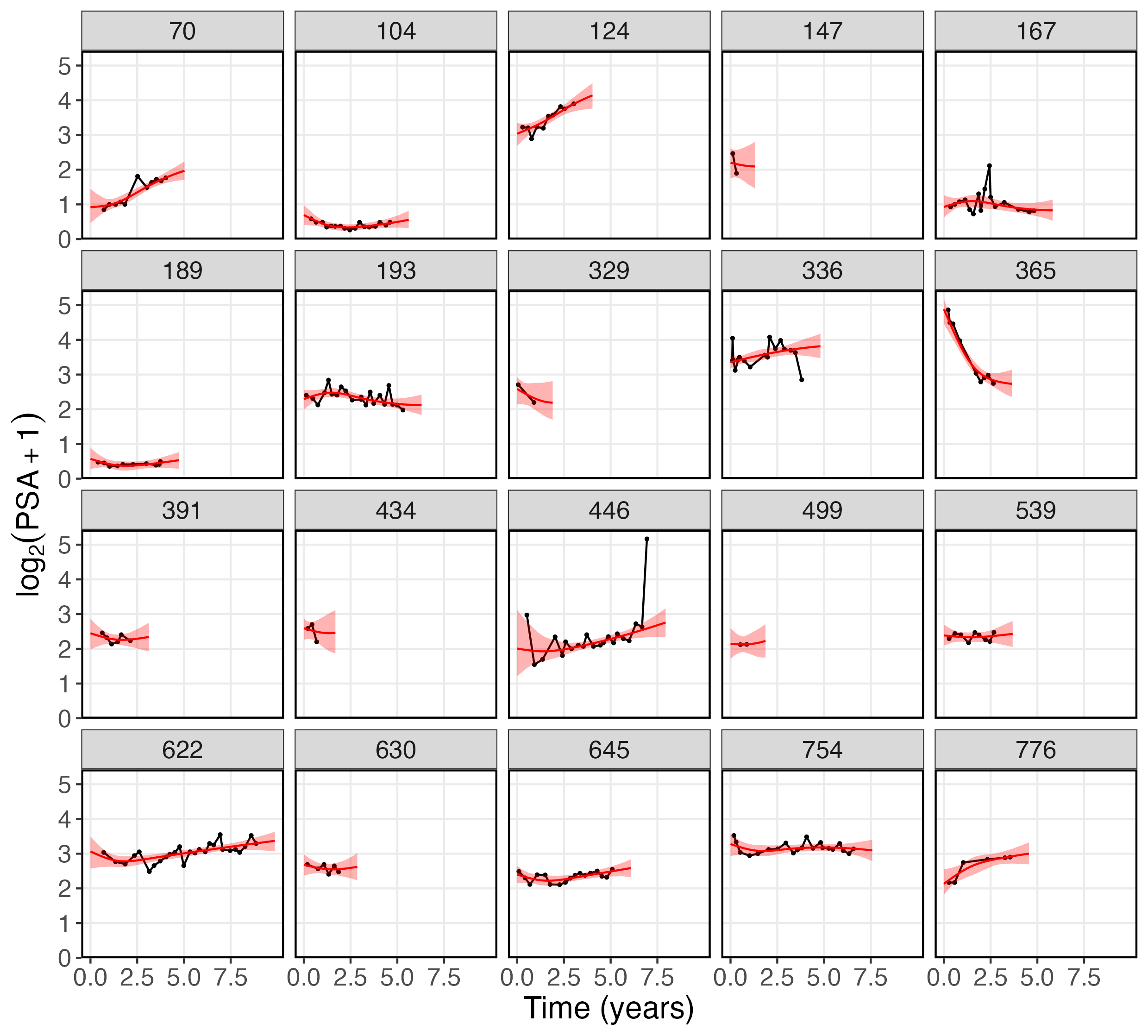}}
    \subfigure[Core ratios]{\includegraphics[width = 0.8\textwidth]{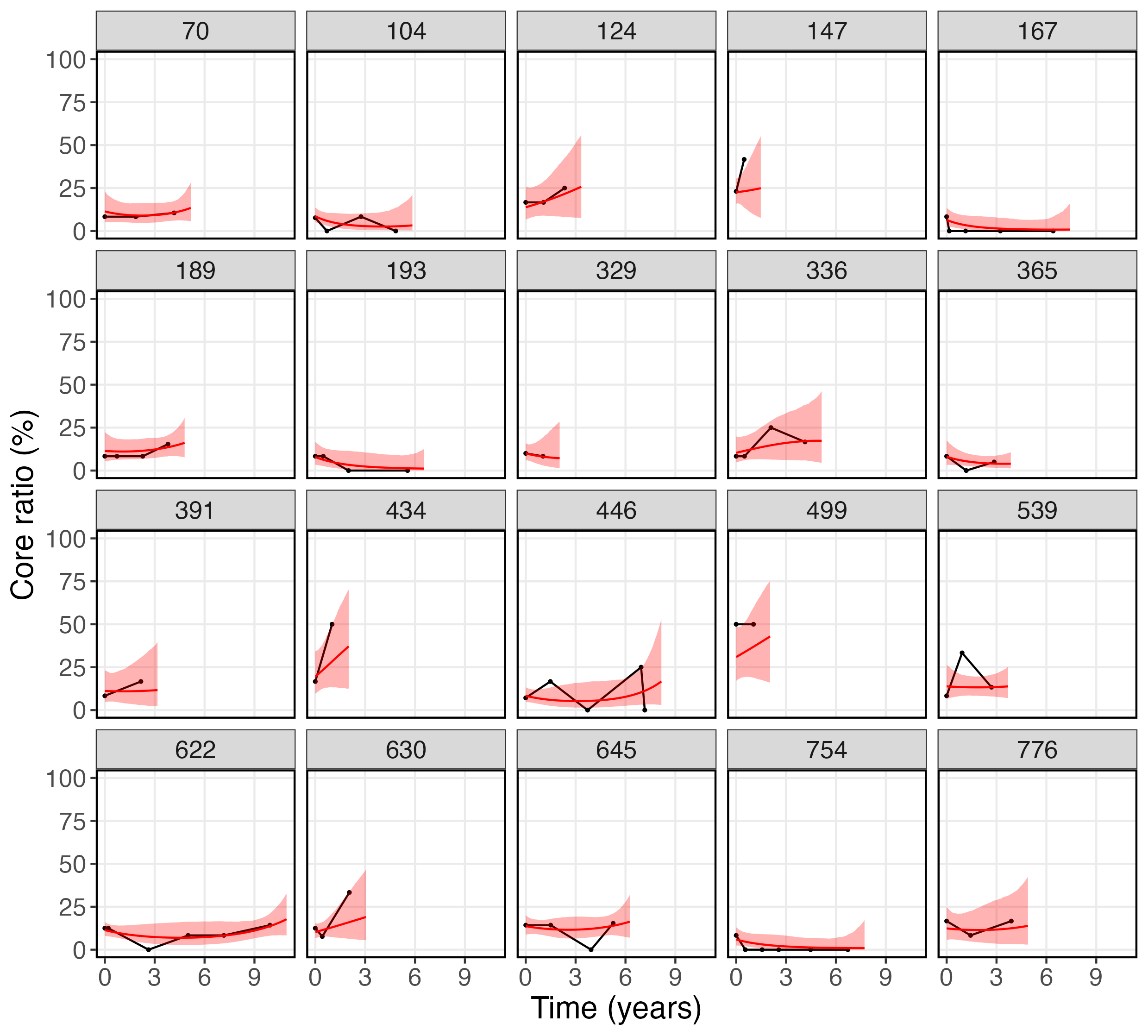}}
    \caption{Fitted trajectories of two longitudinal outcomes for 20 randomly selected subjects.}
    \label{fig:fittedtrajectory}
\end{figure}

The goodness of fit the ICJM depends on the fit of individual expected trajectories of the biomarkers. Therefore, we examined the fitted trajectories of PSA levels and core ratios for the 20 selected subjects in Figure~\ref{fig:observedtrajectory}. The results are visualized in Figure~\ref{fig:fittedtrajectory}. It is shown that the ICJM is able to generally capture the non-linear trends for the biomarkers of each individual patient.

\subsection{Results} \label{subsection:results12}

\begin{figure}[H]
   \centering
   \subfigure[]{\includegraphics[width=0.48\textwidth]{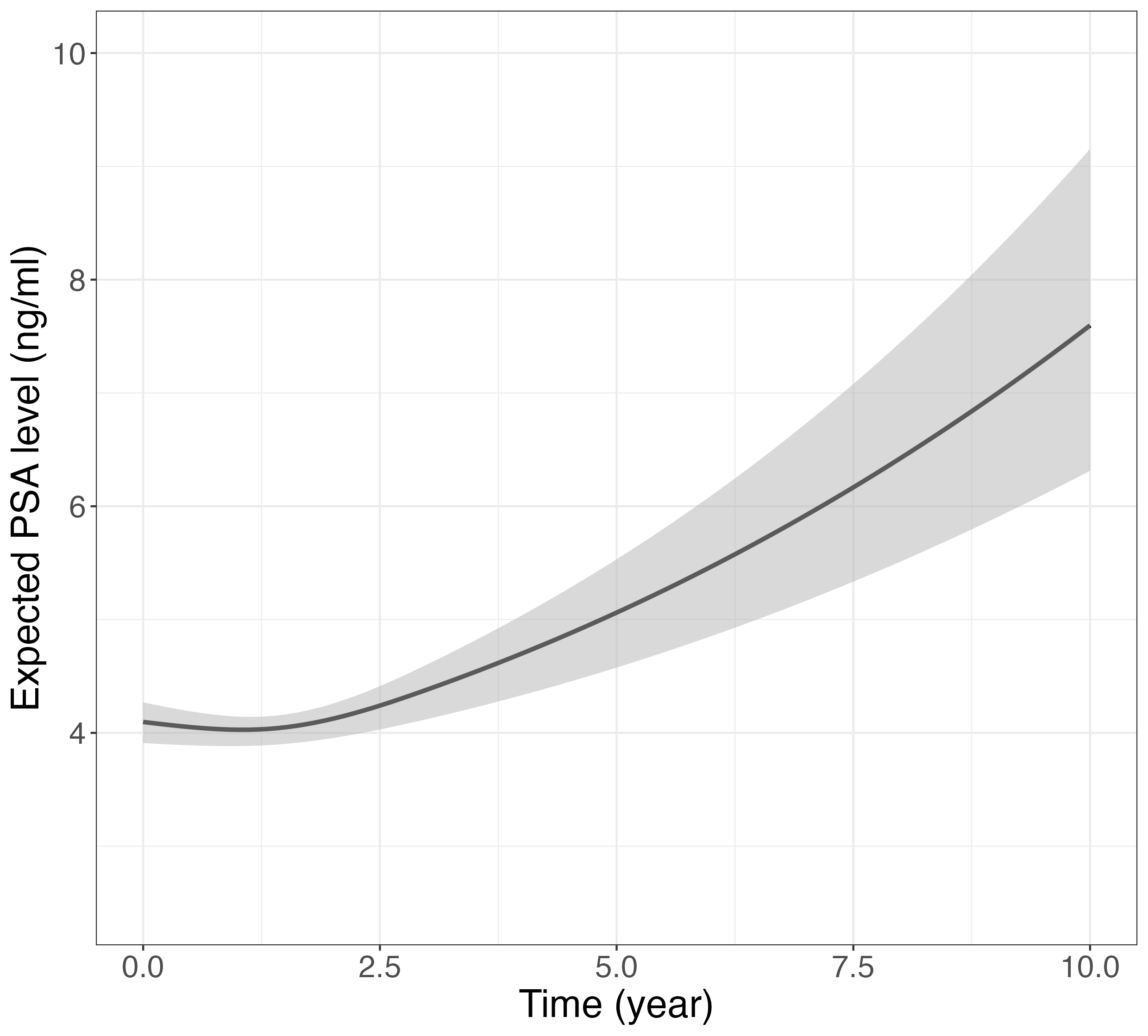}}
   \subfigure[]{\includegraphics[width=0.48\textwidth]{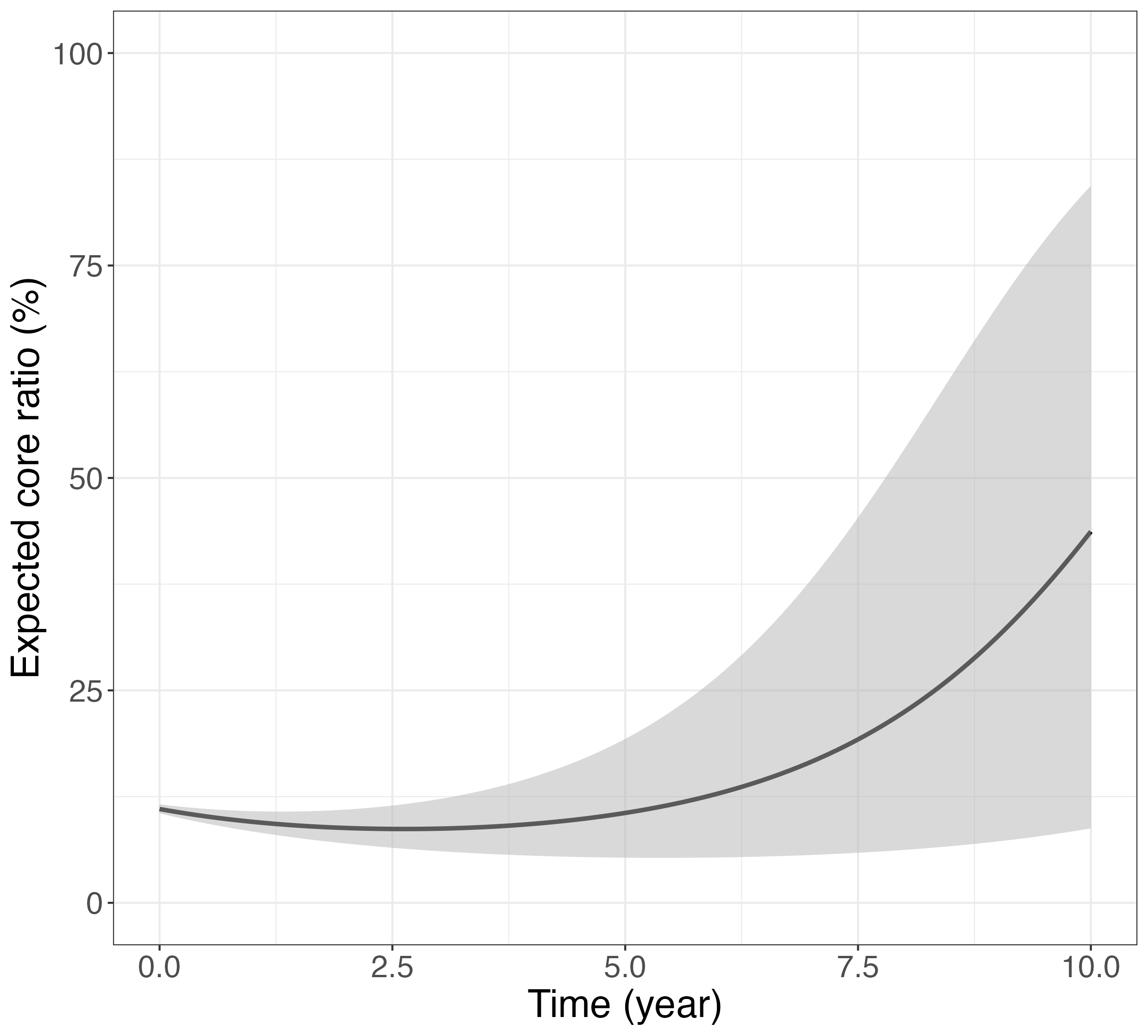}}
   \caption{Effect of time in the longitudinal modeling of (a) the PSA level and (b) the core ratio (ICJM 2).}
   \label{fig:effplottrend}
\end{figure}

The results from ICJM 1 and ICJM 2 are presented in Table~\ref{Tab:ts2}. Since the coefficients pertaining to the natural cubic splines used to model the non-linearity of the trajectories of the PSA levels do not have a direct and clinically meaningful interpretation and to facilitate the interpretation, we present effect plots of the estimated PSA level trajectory and core ratio trajectory for patients with the median age of 62 years in Figure~\ref{fig:effplottrend}. Both outcomes remained stable at the beginning and increased with time.

\begin{table}[H]
    \centering
    \setlength\extrarowheight{3pt}
    \caption{Summary of the model parameter estimates in ICJM 1 and ICJM 2.}
    \begin{tabular}{lcccc}
        \toprule
        \multirow{2}{*}{\textbf{Parameters}} & \multicolumn{2}{c}{\textbf{ICJM 1 (PSA)}} & \multicolumn{2}{c}{\textbf{ICJM 2 (PSA + core ratio)}} \\ \cline{2-5}
         & \textbf{Estimate} & \textbf{95\% CI} & \textbf{Estimate} & \textbf{95\% CI} \\ \midrule
        \multicolumn{5}{l}{\textbf{Longitudinal component - PSA}} \\ 
        Intercept    & 2.35 & [2.30, 2.40] & 2.35 & [2.30, 2.40]  \\
        Time 1$^\dagger$ & 0.28 & [0.19, 0.37] & 0.29 & [0.19, 0.37]  \\
        Time 2$^\dagger$ & 0.59 & [0.41, 0.74] & 0.64 & [0.50, 0.79] \\
        Time 3$^\dagger$ & 0.94 & [0.62, 1.21] & 1.04 & [0.80, 1.32] \\
        Age & 0.02 & [0.01, 0.02] & 0.02 & [0.01, 0.02]  \\ \hline
        \multicolumn{5}{l}{\textbf{Longitudinal component - core ratio}} \\ 
        Intercept    & - & - & -2.09 & [-2.14, -2.03]  \\
        Time & - & - & -0.20 & [-0.28, -0.13]  \\
        Time\textsuperscript{2} & - & - & 0.04 & [0.03, 0.05] \\ \hline
        \multicolumn{5}{l}{\textbf{Progression-specific survival component}} \\ 
        $\log(\text{PSA density})$ & 0.51 & [0.23, 0.77] & 0.30 & [-0.03, 0.58] \\
        $\log_2(\text{PSA} + 1)$ value & 0.13 & [-0.11, 0.36] & 0.26 & [0.02, 0.50] \\ 
        $\log_2(\text{PSA} + 1)$ yearly change & 2.92 & [1.75, 4.10] & 1.65 & [0.30, 2.98] \\
        $\text{logit[E(core ratio)] value}$ & - & - & 1.10 & [0.91, 1.29] \\ \hline
        \multicolumn{5}{l}{\textbf{Treatment-specific survival component}} \\ 
        $\log(\text{PSA density})$ & 0.22 & [-0.18, 0.63] & -0.25 & [-0.79, 0.26]  \\
        $\log_2(\text{PSA} + 1)$ value & 0.42 & [0.08, 0.74] & 0.54 & [0.13, 0.93]\\
        $\log_2(\text{PSA} + 1)$ yearly change & 2.21 & [0.16, 4.21] & 0.51 & [-1.69, 2.82] \\
        $\text{logit[E(core ratio)] value}$ & - & - & 1.54 & [1.20, 1.89] \\ \bottomrule
        \multicolumn{5}{l}{\small $^\dagger$ Time variable is specified using natural cubic splines with 3 degrees of freedom.} \\
        \multicolumn{5}{l}{\small CI: credible interval}
    \end{tabular}
    \label{Tab:ts2}
\end{table}

\begin{figure}[H]
   \centering
   \includegraphics[width=0.8\textwidth]{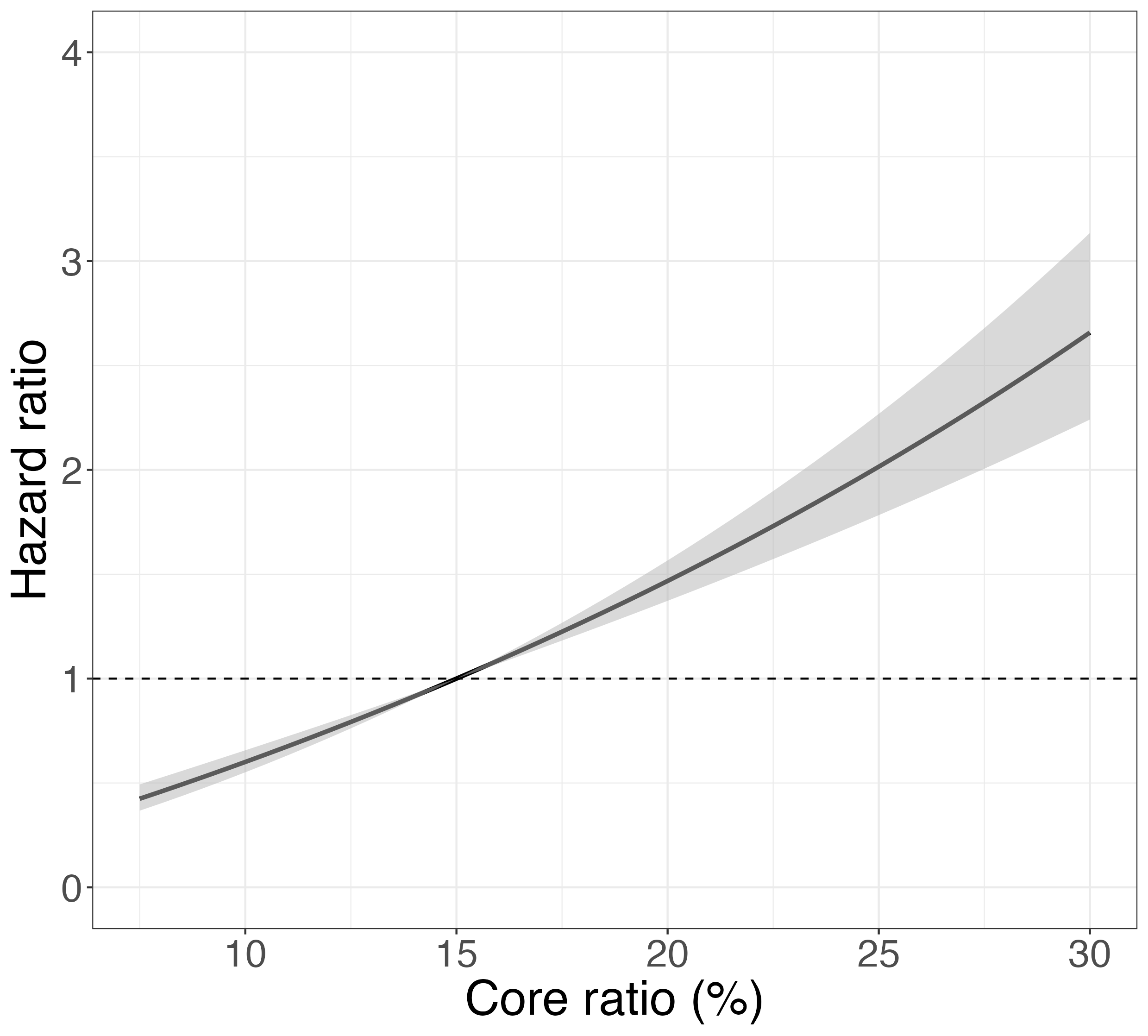}
   \caption{Effect of the core ratio value (contrast to a core ratio of 15\%) on the risk of progression, considering the baseline PSA density, the current PSA level and PSA magnitude of change over the past year remain constant.}
   \label{fig:effplotcrvalue}
\end{figure}

The estimated effect of the core ratio is visualized in Figure~\ref{fig:effplotcrvalue}. A one-fold increase of the core ratio to 30\% raised the progression-specific risk by a factor of 2.66 times while halving the core ratio lowered the risk by 58\%.

\section{Example of a personalized biopsy scheduling procedure} \label{section:example}

In this section, we provide an example to illustrate the dynamic personalized biopsy scheduling methodology.

Imagine patient $i'$ started AS two years ago (i.e., the current time is 2, we currently at visit number $v=4$, i.e., the time of the fourth visit is $t_4=2$) and has just had his regular PSA measurement (at year 2). So far, he has had a biopsy when he entered AS (i.e., $t^{(b)} = 0$), and from this biopsy, the core ratio was determined. So we know the history of the longitudinal outcomes for patient $i'$ up until $t^{(y)} = 2$.

To determine whether a biopsy should be performed at the current visit, we use (3) from the manuscript.
Assume the risk threshold $\phi$ is 0.1 and the calculated progression specific risk $\Pi_{i'}^{(\textsc{prg})}\left\{t_4 \mid t^{(b)}, t^{(y)}\right\} = \Pi_{i'}^{(\textsc{prg})}\left\{2 \mid 0, 2\right\}=0.08$. Since the risk is between the threshold, no biopsy is performed at this time.

To create a schedule of the expected future biopsy times for patient i', we also predict the risk at the subsequent clinical visits, i.e., times $2.5, 3, 3.5, \ldots$.
For predicting the risk at the next (i.e., fifth) visit, at $t_5 = 2.5$, the information on the longitudinal outcomes remains the same since we are still at time 2, and only the time for which we want to estimate the risk has changed. 
Say, the expected risk  $\Pi_{i'}^{(\textsc{prg})}\left\{t_5 \mid t^{(b)}, t^{(y)}\right\} = \Pi_{i'}^{(\textsc{prg})}\left\{2.5 \mid 0, 2\right\} = 0.11$ and, thus, larger than the threshold $\phi = 0.1$.
A biopsy will be scheduled for the visit at 2.5 years.

\begin{figure}[H]
    \centering
    \includegraphics[width = 0.99\textwidth]{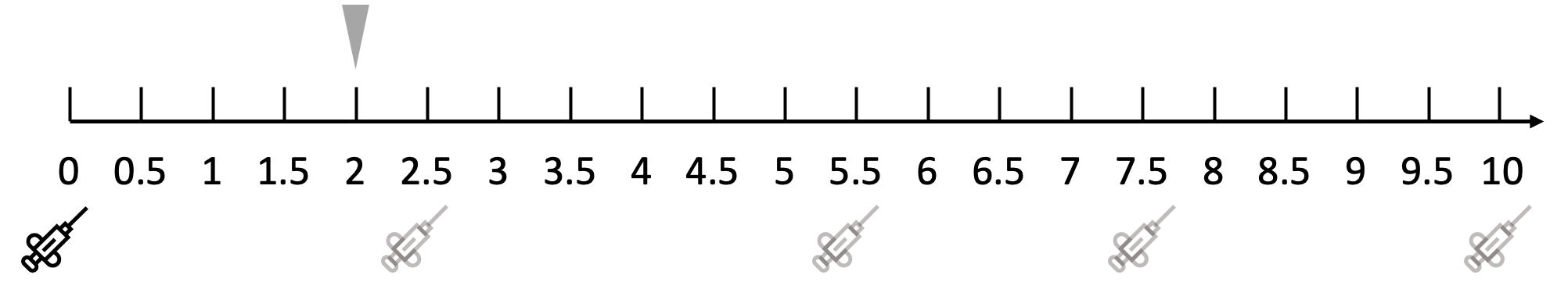}
    \caption{Patient $i'$ at year 2.}
    \label{fig:example1}
\end{figure}

For calculating the expected risk at the subsequent visit times, we assume that this biopsy at the 2.5-year visit will not reveal cancer progression. The risk at year 3 is then calculated conditioning on progression not happening before this most recent (tentative) biopsy at $\tilde{t}^{(b)} = 2.5$. Note that we do not make any assumptions about the yet unknown values of the longitudinal outcomes between the current time and future visits, i.e., we use $\Pi_{i'}^{(\textsc{prg})}\left\{t_7 \mid \tilde{t}^{(b)}, t^{(y)}\right\}$ where $t^{(y)}$ is still the same as before. Following this procedure, we plan biopsies at years $2.5, 5.5, 7.5$, and finally $10$, as visualized in Figure~\ref{fig:example1}.

\begin{figure}[H]
    \centering
    \includegraphics[width = 0.99\textwidth]{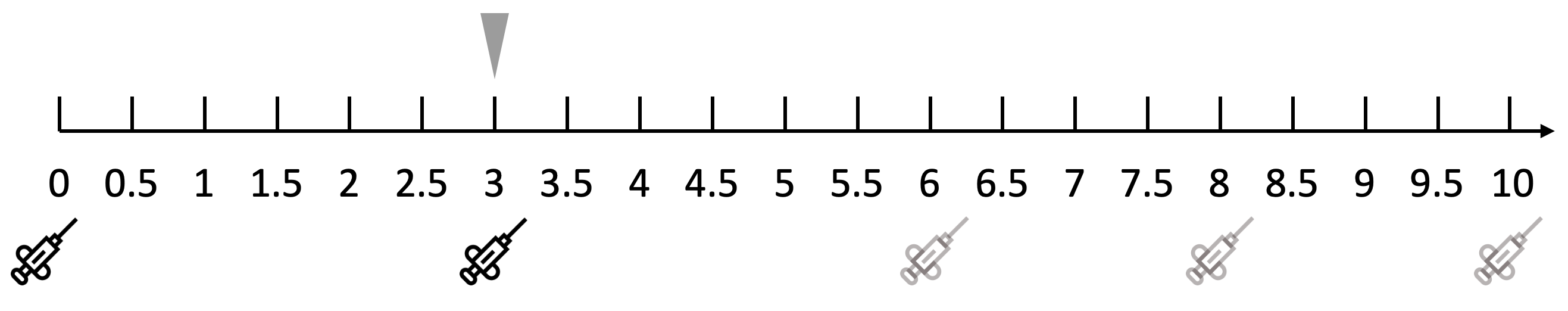}
    \caption{Patient $i'$ at year 3.}
    \label{fig:example2}
\end{figure}

Over time, when additional longitudinal measurements are taken, $t^{(y)}$ is updated as well. Consequently, the personalized schedules are re-generated each time $t^{(y)}$ is updated. Say a half year later, at $t_5=2.5$, a new PSA measurement was taken and the updated risk of progression at this time is $0.09$ and is no longer above the risk threshold. Therefore, a biopsy is not conducted. However, at the following visit $t_6=3$, the again updated risk exceeds the boundary and, thus, a biopsy will be conducted. At this time, a new schedule is proposed, resulting in biopsies being scheduled at years $6, 8,$ and $10$ (Figure~\ref{fig:example2}).

\bibliographystyle{SageV}
\bibliography{myref}